\documentstyle[multicol,prb,aps,graphics]{revtex}
\begin{document}
\title{Electrical resistivity at large temperatures: Saturation and lack
thereof}
\author{M. Calandra and O. Gunnarsson}
\address{ Max-Planck-Institut f\"ur Festk\"orperforschung 
D-70506 Stuttgart, Germany}

\maketitle
\begin{abstract}
Many transition metal compounds show a saturation of the electrical
resistivity at high temperatures, $T$, while the alkali-doped 
fullerenes and the high-$T_c$ cuprates are usually considered to
show no saturation.  We present a model of transition metal compounds, 
which shows saturation, and a model of alkali-doped fullerenes,
which shows no saturation. The electron scattering is assumed
to be due to interaction with phonons. The properties of these 
models are determined by performing quantum Monte-Carlo calculations. 
To analyze the results, as well as earlier results for the high-$T_c$ 
cuprates, we use the f-sum rule. We demonstrate that the f-sum rule    
leads to a natural upper limit for the resistivity at large $T$. 
For some systems and at low $T$, the resistivity increases so 
rapidly  that this upper limit is approached for experimentally
accessible temperatures. The resistivity then saturates.
For a model of transition metal compounds with weakly interacting 
electrons, the upper limit corresponds to an apparent mean free 
path consistent with the Ioffe-Regel condition. For a model of 
the high $T_c$ cuprates with strongly interacting electrons, 
however, the  upper limit is much larger than the Ioffe-Regel 
condition suggests. This upper limit is not exceeded by experimental
resistivities. The experimental data for the cuprates are therefore
consistent with saturation.  After saturation the resistivity 
normally grows slowly. The alkali-doped fullerenes can be 
considered as systems where saturation has happened already for 
$T=0$, due to orientational disorder. We show, however, that for 
these systems the resistivity grows so rapidly after ``saturation'' 
that this concept is meaningless. This is due both to the small 
band width and to the coupling to the level energies of the 
important (intramolecular) phonons in the fullerenes.  
\end{abstract}
\begin{multicols}{2}
\section{Introduction}\label{sec:a}
\begin{figure}
\centerline{
\rotatebox{-90}{\resizebox{!}{3.5in}{\includegraphics{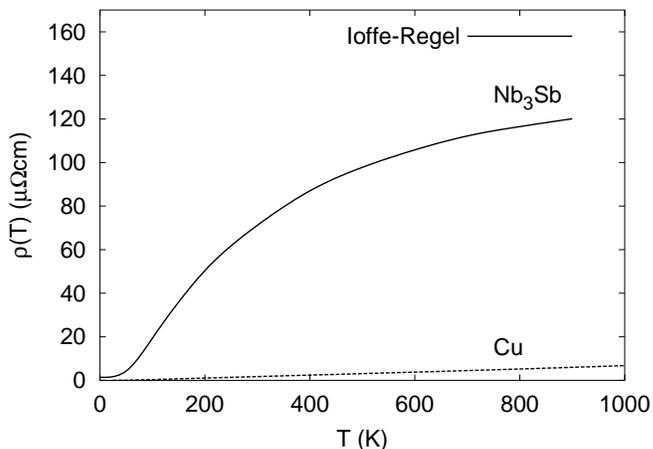}}}}
\caption[]{\label{fig:all0}Resistivities of Cu and Nb$_3$Sb.\cite{Fisk} 
The figure also shows the Ioffe-Regel\cite{ioffe} saturation resistivity
for Nb$_3$Sb, obtained by assuming that the mean free path $l$ in Eq. 
(\ref{eq:i1}) is equal to the the distance between the scattering centers. 
The figure illustrates that for Nb$_3$Sb the resistivity saturates at 
roughly the value expected from the Ioffe-Regel criterion.\cite{ioffe}} 
\end{figure}

The electrical resistivity of metals is often described in a semiclassical
picture, where an electron on the average travels a mean free path $l$
before it is scattered by a phonon, an impurity or another electron.
Assuming a spherical Fermi surface, the resistivity $\rho$ can be expressed
in terms of $l$ as 
\begin{equation}\label{eq:i1}
\rho={3\pi^2 \hbar\over e^2k_F^2 l},
\end{equation}
where $k_F$ is the Fermi wave vector. Alternatively, if we know the
resistivity experimentally, we can deduce an apparent mean free path 
from Eq. (\ref{eq:i1}). For a good metal, $l$ is typically several 
hundred \AA \ or more. As the temperature $T$ is increased, $\rho$
increases. Normally, it is found that $\rho(T)\sim T$ for $T$ 
larger than some fraction of a typical phonon energy. This is 
due to the increased scattering by phonons, and it corresponds 
to a reduction of $l$. Nevertheless, at the melting point, $l$ 
is still typically very much larger than the separation $d$ of 
two neighboring atoms. An example of this behavior is given by Cu 
in Fig. \ref{fig:all0}.

In the 1970's a number of exceptions to this behavior were 
found.\cite{Fisk} In particular for several A15
compounds, such as  Nb$_3$Sb and Nb$_3$Sn, it was found that 
$\rho$ increases very rapidly with $T$ for small $T$, leading to 
very large values already for  temperatures of the order of a few 
hundred K. At these values of $T$, the slope of $\rho(T)$ is strongly    
reduced. This is shown in Fig. \ref{fig:all0}, where the resistivities
of Nb$_3$Sb and Cu are compared. This was described as 
``resistivity saturation''.\cite{Fisk} Interestingly, 
it was found that saturation happened when $l\sim d$, the 
Ioffe-Regel condition.\cite{ioffe} The corresponding resistivity
is also shown in Fig. \ref{fig:all0}.  During the 1970's and 
early 1980's many examples of this were studied, and saturation 
of the resistivity when $l\sim d$ was considered a universal 
behavior.\cite{Allen}

In a semiclassical picture, this behavior may be expected.   
It may seem that the worst that could happen is that an electron
is scattered at every atom. We would then expect $l \sim d$
to be fulfilled. This argument is, however, not convincing.
In the semiclassical theory,
it is assumed that an electron travels through the solid
with a well-defined ${\bf k}$-vector between the scattering 
events. If, however, $l\sim d$, it is not possible to define 
${\bf k}$, and the theory breaks down.\cite{Kohn} A proper theory 
of saturation is therefore needed. A number of theories
have been put forward,\cite{Chakraborty,Cote,Ron} but no 
theory has been generally accepted. Due to the break-down of 
the semiclassical theory when $l\sim d$, the concept of a 
mean free path itself becomes questionable for such small values 
of $l$. In this case we use Eq. (\ref{eq:i1}) as a definition 
of the (apparent) mean free path. 

\begin{figure}
\centerline{
\rotatebox{-90}{\resizebox{!}{3.5in}{\includegraphics{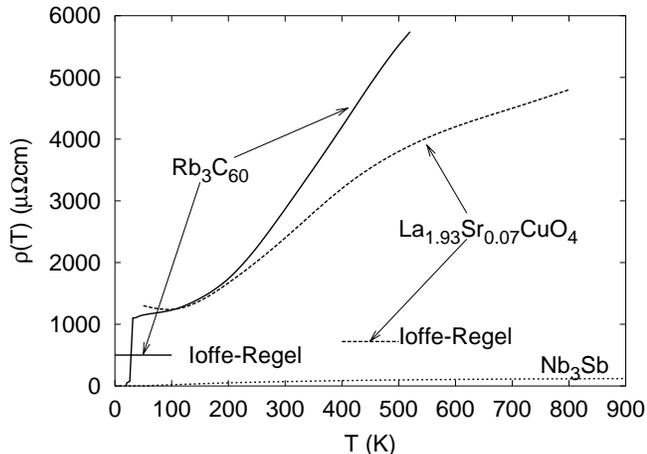}}}}
\caption[]{\label{fig:c601}Resistivities of 
La$_{1.93}$Sr$_{0.07}$CuO$_4$\cite{Takagi} and 
Rb$_3$C$_{60}$\cite{Hebard} and the corresponding 
Ioffe-Regel\cite{ioffe} saturation 
resistivities.\cite{scatteringcenters} The figure illustrates
that the resistivity of these systems becomes much larger 
than predicted by the Ioffe-Regel condition.}
\end{figure}

More recently, several apparent exceptions to resistivity 
saturation have been found. In particular, this is the case for 
some strongly correlated systems, for instance the high-$T_c$ 
cuprates,\cite{Takagi} and for the alkali-doped 
fullerenes.\cite{Hebard,Housat} This is illustrated 
in Fig \ref{fig:c601}, where we show the resistivities of
La$_{1.93}$Sr$_{0.07}$CuO$_4$ and Rb$_3$C$_{60}$ together 
with the Ioffe-Regel resistivities.\cite{scatteringcenters}
Different experiments for alkali-doped C$_{60}$ compounds
show substantial differences, but this is not essential for the 
present discussion. The Ioffe-Regel resistivities of these two 
systems are very large, due to the low carrier density. The 
figure illustrates that the experimental resistivities,nevertheless, 
greatly exceed the Ioffe-Regel resistivities. 
It also illustrates that the resistivities of these two compounds
are very much larger than for Nb$_3$Sb and other systems, which shows
saturation according to the Ioffe-Regel condition. 

This shows that the semiclassical argument behind the Ioffe-Regel 
condition is not only questionable, but that it leads to wrong
conclusions for the high-$T_c$ cuprates and the C$_{60}$ compounds.
This emphasizes the need for a proper theory of why saturation 
happens for some systems but not for others. We also need to 
understand why saturation happens for most transition metal 
compounds when $l\sim d$, although $l$ is not a well-defined 
concept any more. 

We have earlier presented such a theory for transition metal 
compounds in a short publication,\cite{saturationprl} and we here 
expand the arguments.  We have also analyzed the reasons for the 
lack of saturation in the alkali-doped C$_{60}$ compounds,\cite{Han} 
and we provide additional results here. Finally, we have also presented 
results for a model of the high-$T_c$ cuprates.\cite{saturationhightc}  
We have therefore considered models of three classes of systems: 
i) a model of weakly correlated transition metal compounds, which 
shows saturation in agreement with the Ioffe-Regel condition, 
ii) a model of strongly correlated high-$T_c$ compounds, which shows 
saturation but at much larger values than predicted by the Ioffe-Regel 
conditions, and iii) a model of alkali-doped fullerenes, which shows 
no saturation. 

We assume that in case i) and iii) the important scattering is 
due to the electron-phonon interaction. In a model Hamiltonian 
approach, there are two natural types of coupling to the phonons, 
either via the level energies (LE coupling) or via the 
hopping matrix integrals (HI coupling). In most nonionic compounds 
the latter effect should be the dominating one. As the 
distance between two neighboring atoms is changed due to 
the excitation of a phonon, the main effect should be a 
change of the hopping integrals.  We study this for a 
model of transition metal compounds, referred to as the 
TM model.

In molecular solids, such as the alkali-doped fullerenes, the
situation is different. Due to the weak coupling between the
molecules, it is sensible to first calculate the levels of a free
molecule, and then to study the weak hopping between these levels.
In the alkali-doped fullerenes the main coupling is to intramolecular
phonons. These phonons couple primarily to the level energies  
and only weakly to the hopping integrals between the molecules. 
We therefore study the LE coupling for a model of alkali-doped 
C$_{60}$ systems, in the following referred to as the C$_{60}$ model. 
The LE coupling may also become important for strongly ionic
systems.

We use a quantum Monte-Carlo (QMC) method\cite{Scalapino} 
for calculating the current-current correlation function for 
imaginary times. A maximum entropy 
method\cite{Jarrell} is then applied to analytically continue 
the response function to the real frequency axis. This gives the 
frequency dependent optical conductivity $\sigma(\omega)$, and  
the resistivity $\rho=1/\sigma(\omega=0)$. Since the QMC method has 
no sign problem for the models studied here, we are able to obtain 
rather accurate results for the resistivity. In particular, we can 
establish whether or not the models we consider show resistivity 
saturation. 

To interpret the results we use a simplified approximate 
approach, treating the phonons (semi)classically. 
By comparing with In this method we assume that the phonons 
can be described by random static displacements of the atoms 
with an average amplitude that increases with $T$. The remaining 
electronic problem can then easily be solved quantum mechanically. 
This approach is in contrast to the Boltzmann equation, where 
the electrons are treated semiclassically. The main advantage of 
this method, compared with the QMC calculation, is that it is 
simple enough to allow an interpretation of the results. By 
comparing with the QMC results we establish the range of 
applicability of the semiclassical method for the models of 
interest here.  

In our semiclassical treatment, the 
excitation of phonons leads to a static variation of the level energies  
in the C$_{60}$ model and of the hopping integrals in the TM model. 
In the context of disordered systems, this is referred to as diagonal 
and off-diagonal disorder, respectively. Past work has primarily
studied diagonal disorder, which in some respects is technically
simpler.

Traditionally, transport is described within the Boltzmann theory. 
The Bloch-Boltzmann\cite{Bloch} theory starts from the perfectly 
periodic system, and treats the scattering mechanisms as small 
perturbations. This can be considered as a theory which is valid 
to lowest order in $1/(k_Fl)$.\cite{Kohn} This further empahsizes 
that the Boltzmann equation becomes questionable when $l\sim d$. 
Furthermore, the Ziman approximation\cite{Grimvall} to the 
Boltzmann equation leads to $\rho(T)\sim T$ for large $T$, i.e., 
there is no saturation in contrast to what is found experimentally 
for many systems. It is then natural to look for extensions of the 
Boltzmann equations, which would extend the range of perturbation 
strengths that can be treated.\cite{Chakraborty,Cote} We find, 
however, that in, e.g., the A15 compounds the thermally excited 
phonons even at relatively small $T$ tend to largely remove the 
effects of periodicity.   In the semiclassical treatment of the 
phonons, the momentum conservation in the electronic system is 
lost already for temperatures of the order of a few hundred K. 
We therefore consider the opposite limit to the Boltzmann equation, 
where we assume that thermal excitations have completely destroyed 
periodicity. At low $T$ there is a Drude peak in the optical 
conductivity $\sigma(\omega)$ due to intraband transitions between 
states with similar ${\bf k}$-vectors. As $T$ is increased,    
${\bf k}$-conservation is lost, and the meaning of intraband
transitions is blurred, the Drude peak disappears. We therefore 
focus on the limit where there is no pronounced structure in 
$\sigma(\omega)$ at small $\omega$. 

We have earlier used current and charge conservation to obtain  
simple upper estimates for the resistivity of a metal.\cite{saturationprl}
Here we show how the same result can be derived by using the 
(related) f-sum rule. This approach has the advantage that 
it can also be used to discuss the high-$T_c$ 
cuprates,\cite{saturationhightc} and that it is convenient
for discussing the fullerenes. The approach based on the f-sum 
rule therefore provides the most convenient framework for analyzing 
the different classes of materials. 

We combine the f-sum rule with the assumption that the Drude peak 
is lost.  This naturally leads to an upper limit for the resistivity 
at small or intermediate $T$'s. If the initial slope of $\rho(T)$ 
is very large, $\rho(T)$ reaches this limiting value already for
experimentally available values of $T$. At this point saturation 
normally happens, as is illustrated in our TM model. The removal 
of the Drude peak could be due to any scattering mechanism, e.g., 
electron-phonon (HI or LE coupling), electron-electron or disorder 
scattering. For the TM model considered here, we show in a quantum
mechanical treatment that saturation should happen roughly when the 
Ioffe-Regel criterion is satisfied. This is somewhat accidental and 
it is not true for a model of the high-$T_c$ cuprates, where strong 
correlation effects leads to a larger saturation resistivity. 

While a pronounced saturation is observed for 
the A15 compounds Nb$_3$Sb or Nb$_3$Sn, other systems, such as Nb,
show a weaker saturation or no saturation at all. Here we study
a simple model of A15 compounds, referred to as the Nb$_3^{\ast}$ 
model,\cite{Pickett}  where we include the $d$-orbitals of the Nb atoms,
put on the appropriate A15 lattice, but where the remaining atom 
(e.g., Sn in Nb$_3$Sn) is neglected. This is compared with Nb, where 
the atoms are put on a bcc lattice. These two models then only differ 
with respect to the lattice structure. This difference leads to
a smaller plasma frequency for Nb$_3$Sb and a steeper slope
of $\rho(T)$. This leads to a much more pronounced saturation 
for Nb$_3$Sb. 

Even after ``saturation'' has happened, $\rho(T)$ tends 
to continue to grow, but at a slower rate. In this respect
there is sometimes an essential distinction between LE and HI 
coupling.  This can be best discussed using the f-sum
rule. We show that the change of the resistivity can be viewed  
as resulting from a change of the kinetic energy and of the band 
width. These changes keep growing without limit with $T$, due 
to the Bose nature of the phonons and the lack of limitation
on the number of phonons. The two changes work together for the 
LE coupling, but tend to compensate each other for the HI coupling. 
As a result the resistivity grows more slowly after 
``saturation'' for the HI coupling and the saturation is 
more pronounced. This distinction is fairly clear-cut for the 
C$_{60}$ model. For this model, disorder leads to
such a strong scattering, that ``saturation'' can be considered
to have happened already at $T=0$. Due to the LE coupling and 
the small band width, however, $\rho(T)$ grows so rapidly after 
``saturation'' that the concept of saturation becomes meaningless. 
For HI coupling, on the other hand, the resistivity shows a clear
change in slope, even  for the C$_{60}$ model.

In Sec. II we present the TM and C$_{60}$ models and in Sec. III
the QMC and semiclassical methods are described. The results
are presented in Sec. IV and discussed in Sec. V.
In Sec. VI we summarize the present results as well as earlier results
for the High $T_c$ cuprates in the framework of the f-sum rule.

\section{Models}\label{sec:b}

\subsection{TM model}\label{sec:btm}

We first consider a model appropriate for a transition metal 
(compound), referred to as the TM model.  Each transition
metal atom has a five-fold degenerate ($n=5$) level. It couples 
to the other atoms via hopping matrix elements $t_{\mu\nu}$, where
$\nu\equiv (m,i)$ is a combined label for a orbital index $m$ and 
a site index $i$. Thus the electronic Hamiltonian is 
\begin{equation}\label{eq:m1}
H^{\rm el}=\varepsilon_0\sum_{\mu\sigma}c^{\dagger}_{\mu\sigma}c_{\mu\sigma}
+\sum_{\mu\nu\sigma}t_{\mu\nu} c^{\dagger}_{\mu\sigma}c_{\nu\sigma},
\end{equation}
where $c^{\dagger}_{\mu}$ creates an electron in the state 
$|\mu\rangle$. As discussed in the introduction, we consider 
two different models where the atoms are put on a bcc or an A15
lattice, describing a transition metal (Nb) or an A15 compound, 
respectively. As discussed above, in the case of the A15 compound
we only consider the transition metal atoms and, for instance, 
neglect Sb in Nb$_3$Sb.\cite{Pickett} This is referred to as the 
Nb$_3^{\ast}$ model. 
 
To describe the  hopping integrals, we essentially follow
Harrison,\cite{Harrison} and assume that the radial part
of the integrals has  a power dependence on the separation
of the atoms. However, instead
of the power five, used by Harrison, we use the power 3.6, more
appropriate for Nb.\cite{ove}
Using Harrison notation for the radial part between two atomic
$d$ energy levels,
\begin{equation}
\label{eq:m2}
V_{dd,s}=\eta_{dd,s} \frac{\hbar^2 r_d^{1.6}}{m}\frac{1}{ |{\bf R}_i-{\bf R}_j|^
{3.6}+a_0^{3.6}}
\end{equation}
where $\eta_{dd,\sigma}=-16.2$,$\,\eta_{dd,\pi}=8.75$ and $\eta_{dd,\delta}=0$
and $m$ is the electron mass.
The parameter $r_d$ has been chosen in order to reproduce the band with as
obtained from LDA calculations for $Nb_3^*$\cite{Pickett}, namely $r_d=0.7$.
Since the atoms vibrate, their separation can occasionally become
very small. To avoid that the hopping integrals then become 
very large, we have introduced the term containing $a_0$ in the
denominator. We use $a_0=2$ \AA. Eq. (\ref{eq:m2}) shows the
distance dependence. In addition there are angular factors, depending
on which $m$-quantum numbers are involved, as described by 
Harrison.\cite{Harrison} In the model of Nb we only consider nearest
neighbor hopping, while in the A15 model (Nb$_3^{\ast}$) also second 
nearest neighbor hopping is included, since the second nearest neighbors 
are not much further away then the nearest neighbors. 

We consider the case when the phonons couple to the hopping integrals
(HI). The phonons are approximated as Einstein phonons. The frequency
$\omega_{ph}=0.014$ eV was obtained from the average frequency 
of Nb metal.\cite{Wolf} For each Nb atom we introduce one such phonon
in each coordinate direction. The $x$-coordinate of atom $i$ is then
given by
\begin{equation}\label{eq:m3}
R_{ix}=R^0_{ix}+\sqrt{\hbar \over 2 M\omega_{ph}}
(b_{ix}+b_{ix}^{\dagger}),
\end{equation}
where $R_{ix}^0$ is the unperturbed $x$-coordinate of the 
atom $i$, $b_{ix}^{\dagger}$ creates a phonon in the $x$-direction 
on site $i$ and $M$ is the mass of a Nb atom. These vibrations 
couple to the hopping matrix elements.  

To obtain the conductivity we calculate the current-current 
correlation function. This requires a definition of the matrix elements 
of the current operator. In our model Hamiltonian approach, it is not
appropriate to calculate these as expectation values of the current
operator between some basis functions, since the basis functions
underlying our model Hamiltonian are not explicitly defined. Instead
one can use charge and current conservation, i.e., the requirement 
that the change of density inside some small volume is equal to the 
current entering this volume. This leads to the result
\begin{equation}\label{eq:m4}
{\bf \hat j}_{\mu\nu}={ie\over \hbar}({\bf R}_i-{\bf R}_j)t_{\mu\nu},
\end{equation}
where $\mu\equiv (m,i)$ and $\nu\equiv (m^{'},j)$.

\subsection{C$_{60}$ model}\label{sec:bc}

We next consider a model appropriate for alkali-doped fullerenes,
referred to as the C$_{60}$ model.
In these systems the $t_{1u}$ band is partly occupied, and we 
therefore consider a model with a three-fold degenerate $t_{1u}$ 
orbital on each C$_{60}$ 
molecule $i$. These orbitals are connected by nearest neighbor 
hopping matrix elements. For the electronic part we therefore 
use the same form of the Hamiltonian as above (\ref{eq:m1}),
but the orbitals are now three-fold degenerate and placed on 
a fcc lattice.  

The hopping integrals are obtained from a tight-binding
description.\cite{Orientation,Satpathy} For each of the 60 C atoms 
in a C$_{60}$ molecule we introduce one $2p$ orbital pointing 
radially out from the molecule.  We then generate orbitals of 
$t_{1u}$ character by forming a linear combination of the 60 $2p$ 
orbitals. The hopping between the $t_{1u}$ orbitals on different 
molecules is then determined by the hopping between $2p$ orbitals
on different molecules. The $2p$ orbitals couple via $\sigma$ and 
$\pi$ hopping integrals.  We use
\begin{eqnarray}\label{eq:dist}
&&V_{\sigma}=V_0 de^{-(d-d_0)/L}  \\
&&V_{\pi}=-V_{\sigma}/4
\end{eqnarray}
where $V_0=9.85$ eV, $d_0=1.43$ \AA \ and $L=0.505$ \AA. 
The calculations
were performed for the lattice parameter 14.24 \AA. In most calculations
we take into account\cite{Orientation,MazinAF} the orientational 
disorder\cite{Stephens} of the C$_{60}$ molecules.

The important electron-phonon coupling is due to the intramolecular 
phonons of H$_g$ symmetry. There are eight such phonons in C$_{60}$,
each one being a five-fold degenerate Jahn-Teller mode. Here we only 
include one degenerate H$_g$ mode per site. We use the Hamiltonian 
\begin{eqnarray}\label{eq:m6}
&&H^{\rm el-ph}= \\ 
&&+ {g\over 2} \sqrt{{2M\omega_{ph}}\over \hbar}
\sum_{\gamma=1}^5 \sum_{i\sigma}\sum_{m=1}^3
\sum_{m^{'}=1}^3
V_{mm^{'}}^{(\gamma)}\psi^{\dagger}_{im\sigma} \psi_{im^{'}\sigma}
x_{i\gamma}, \nonumber
\end{eqnarray}
where $x_{i\gamma}$ is the phonon coordinate for           
a phonon with quantum number $\gamma$ on site $i$, $g$ is an overall 
coupling strength and $V^{(\gamma)}_{mm^{'}}$ are dimensionless coupling 
constants\cite{Lannoo,c60jt} given by symmetry. The dimensionless
electron-phonon coupling constant is given by
\begin{equation}\label{eq:m7}
\lambda=5{g^2\over \omega_{ph}}N(\mu),
\end{equation}
where $N(\mu)$ is the density of states per spin, orbital and 
molecule at the Fermi energy. The current matrix elements are 
given by Eq. (\ref{eq:m4}) with ${\bf R}_i={\bf R}_i^0$.

As a comparison, we also consider a C$_{60}$ model where the 
intermolecular phonons couple to the hopping integrals (HI coupling),
instead of the LE coupling considered above. This coupling is 
obtained by displacing the molecules  
from their ideal positions of the fcc lattice due to the excitations 
of intermolecular phonons. For large values of $T$, the molecules 
come unrealistically close to each others in our semiclassical
theory, neglecting the strongly repulsive interaction for
small separations, and the hopping integrals become 
unrealistically large. For this reason we introduce a modification 
of the hopping integrals between the $2p$ orbitals in the case 
of the HI coupling . The exponent
$e^{-(d-d0)/L}$ is replaced by
\begin{equation}\label{eq:cutoff}
e^{-(d_1-d_0)/L}{  e^{(d_1-d0)/L}+e^{(d_2-d_0)/L}\over 
e^{(d-d0)/L}+ e^{(d_2-d_0)/L}}
\end{equation}
where $d_1=3.1$ \AA \ is the separation of the nearest C atoms 
on neighboring molecules in the equilibrium position and $d_2=2$ 
\AA. For $d\gg d_2$, the hopping integrals are essentially
unchanged, and for $d=d_1$ and they exactly unchanged, while for 
$d\ll d_2$ the hopping integral is cut off at a value which  is 
factor 10 larger than in equilibrium.

\section{Methods}\label{sec:c}

\subsection{Quantum Monte-Carlo method}\label{sec:cqmc}    

To establish the properties of our models, we use a quantum 
Monte-Carlo (QMC) approach.\cite{Scalapino} For these models, 
the QMC method has no so-called sign problem, thanks to the absence
of a repulsive Coulomb interaction. In the calculation of       
response functions for imaginary times there are then only
statistical errors which can be made arbitrarily small by 
improving the sampling. These response functions are analytically 
continued to the real frequency axis by using a maximum entropy 
method.\cite{Jarrell} Although it is nontrivial to control the errors
in this method, it should still be quite accurate for the 
response functions considered here, due to the simple form of their 
spectra.  Thus we are able to quite accurately establish the large $T$
behavior of the resistivity for models with coupling to phonons. 

In the QMC approach used here,\cite{Scalapino} the starting point is
the partition function
\begin{equation}\label{eq:me1}
Z={\rm Tr} e^{-H/T},
\end{equation}
where Tr is a trace over all states. An imaginary time $\tau$ is introduced,
$0\le \tau\le \beta=1/(k_B T)$.  The partition function can then be expressed
as a functional integral over the phonon coordinates as a function of 
$\tau$. For given values of the phonon coordinates, the electronic
part of the Hamiltonian is a one-particle Hamiltonian. The electronic 
degrees of freedom can then be integrated out and be expressed as a 
determinant. Finally, the phonon coordinates are sampled in a Monte 
Carlo approach. 

For the LE coupling, the phonons are local and only influence 
the levels on the molecule of the phonon. For the C$_{60}$ model,
this corresponds to a $3\times 3$ block in the determinant obtained 
in the approach above. The change of the determinant when one phonon 
coordinate is changed can then easily be obtained in an updating 
approach.\cite{Scalapino} For the HI coupling, on the other hand,
each phonon influences the hopping integrals to the neighbors
of the atom of the phonon. Different phonons then couple to partly
``overlapping'' blocks. It is then not possible to introduce the simple
block form used in the C$_{60}$ model. This leads to a substantially
more complicated updating approach, which is discussed in appendix
\ref{sec:TM}.

\subsection{Semiclassical method}\label{sec:cs}
While the QMC method above is very useful in establishing the properties
of our models, its complexity means that it is hard to interpret
the results. We therefore introduce a much simpler method, where the 
phonons are treated semiclassically. We demonstrate that this method 
is quite accurate for the TM model with HI coupling, by showing that
it agrees quite well with the accurate QMC calculations. 
For the C$_{60}$ model with LE coupling, the accuracy is less good,
in particular for large $T$. The method is, nevertheless,
useful for the interpretation. 

We consider a large super cell with $L$ unit cells, $K$ atoms
per unit cell and a total of $N=KL$ atoms. Periodic boundary 
conditions are used. Each phonon coordinate is given a random 
displacement according to a Gaussian distribution centered at zero and
the width
\begin{equation}\label{eq:me2}
\langle x^2 \rangle = {\hbar \over M\omega_{ph}}n_B(T)
\end{equation}
where 
\begin{equation}\label{eq:me2a}
n_B(T)={1\over e^{\hbar \omega_{ph}/(k_B T)}-1},
\end{equation}
is the occupation of the phonon mode.
In this way, a set of displaced coordinates are obtained. These 
define a one-particle Hamiltonian for the electrons. In the case 
of HI coupling, we simply calculate the hopping matrix elements using
the displaced atomic positions. For the LE coupling, we insert
the phonon displacements in Eq. (\ref{eq:m6}). Since the coupling 
contains a factor $\sqrt{M}$, the Hamiltonian is independent of
$M$ for a given $\lambda$ and $\omega_{ph}$ in the case of the
LE coupling. 

To calculate optical conductivity, we find the eigenstates
$|l\rangle$ and eigenvalues $\varepsilon_l$ of this Hamiltonian.
The optical conductivity is then given by
\begin{equation}\label{eq:me3}
\sigma(\omega)={2\pi\over N\Omega \omega}\sum_{ll^{'}}|\langle l|\hat j_x|
l^{'}\rangle |^2(f_l-f_{l^{'}})\delta(\hbar \omega -\varepsilon_{l^{'}}
+\varepsilon_l),
\end{equation}
where $\Omega$ is the volume per atom and $f_l$ is the Fermi 
function for the energy $\varepsilon_l$. The prefactor  two 
comes from the summation over spin. We have assumed that the 
system is isotropic, so that it is no limitation to consider 
the conductivity in the $x$-direction.  

\begin{figure}
\centerline{
\rotatebox{0}{\resizebox{!}{2.5in}{\includegraphics{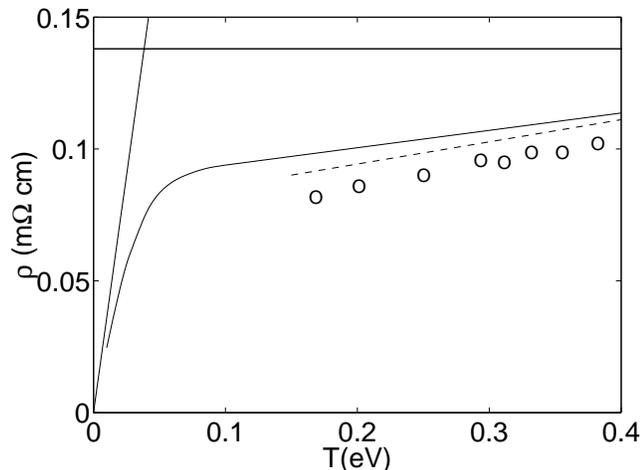}}}}
\caption[]{\label{fig:Nb3}Resistivity $\rho(T)$ as a function of
temperature $T$ for Nb$_3^{\ast}$. The figure compares the 
semiclassical (broken ($N=36$) and full ($N=648$) curves) and  
QMC (circles, $N=36$) calculations. The figure also shows the 
small (Eq. (\ref{eq:d15})) and large (Eq. (\ref{eq:f4})) temperature 
results. The figure illustrates that the resistivity of the TM model 
saturates at large $T$. Comparison with the QMC results, shows that
the semiclassical calculation is quite accurate, at least for large $T$.}
\end{figure}

Fig. \ref{fig:Nb3} compares the QMC (circles) and semiclassical 
(broken curve) methods for Nb$_3^{\ast}$ with $N=36$ atoms in the 
super cell. The QMC calculation has been limited to rather 
large values of $T$, which is the range of particular interest
here, and which is also the range of $T$ where the calculation
can be performed with a reasonable numerical effort. The figure
illustrates that the semiclassical calculation is quite accurate 
at large $T$ for the TM model. By comparing the semiclassical calculation
for $N=36$ and $N=648$ we also illustrate that at large $T$ the 
result does not change much if the size of the super cell is increased.
For small values of $T$, however, the discreteness of the levels for
$N=36$ would prevent a reliable semiclassical calculation for this 
super cell size.

\begin{figure}
\centerline{
\rotatebox{-90}{\resizebox{!}{3.5in}{\includegraphics{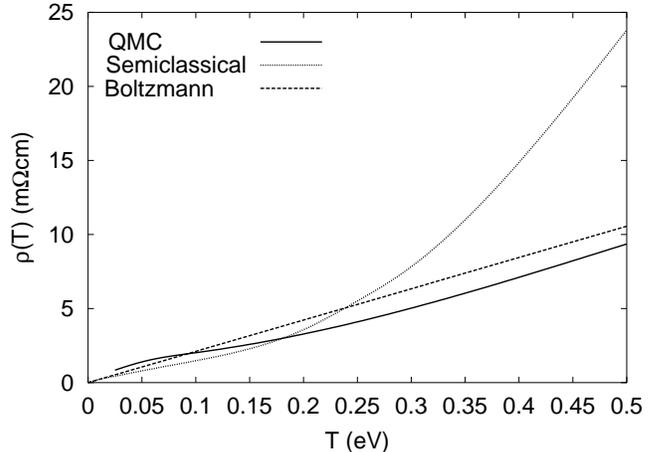}}}}
\caption[]{\label{fig:rhoordlarge}Resistivity $\rho(T)$ as a function of
temperature $T$ in the ordered C$_{60}$ model for $\omega_{ph}=0.00001$ 
eV and $\lambda=0.6$. The figure compares the QMC (full curve),
the semiclassical (dotted curve) and the Boltzmann (broken curve)
results. The phonon frequency was chosen to be so small that $\rho(T)\sim T$
in the Boltzmann theory for all $T$ of interest. }
\end{figure}

Fig. \ref{fig:rhoordlarge} compares the semiclassical theory
(dotted curve) with the QMC (full curve) and the Boltzmann
(broken curve) theories for the C$_{60}$ model with LE coupling, 
assuming ordered C$_{60}$ molecules. The small $T$ behavior is discussed
in detail in Sec. \ref{sec:df}.
Here we just notice that the semiclassical theory agrees with
the Boltzmann theory for very small $T$ and that it agrees 
approximately with the QMC results for small and intermediate 
values of $T$. There is, however, a qualitative disagreement
for large $T$. The reason is that the strong static diagonal disorder 
introduced by the phonons in the semiclassical theory for large
$T$ leads to localization. This is discussed in more detail
in Sec. \ref{sec:fb}. While the semiclassical theory for the
C$_{60}$ model with LE coupling is sufficiently accurate to 
analyze the results  for small and intermediate values of $T$, 
it is less accurate than for the TM model with HI coupling, 
in particular for large $T$. This is further discussed in Sec. 
\ref{sec:fb}.

\section{Results}\label{sec:e}

\subsection{TM model}\label{sec:etm}

The full curve in Fig. \ref{fig:Nb3} shows the semiclassical results 
for the Nb$_3^{\ast}$ model. It illustrates how the resistivity shows 
a very pronounced saturation already at quite small temperatures. 
The calculated resistivity at large $T$ agrees rather well with the 
experimentally results, e.g., about 0.12 m$\Omega$cm at $T=900$ K 
(0.08 eV).\cite{Fisk} This agreement with experiment is important, since,
as we discuss below, our saturation resistivity (Eq. (\ref{eq:f4}))
essentially only depends on the nearest neighbor distance, the 
orbital degeneracy $n$, the filling and the lattice structure. This 
illustrates that our TM model is appropriate for describing resistivity 
saturation. For small $T$, the resistivity grows slower than what 
is found experimentally, which is probably due to the electron-phonon 
interaction being somewhat underestimated in our simple model.

\begin{figure}
\centerline{
\rotatebox{0}{\resizebox{!}{2.5in}{\includegraphics{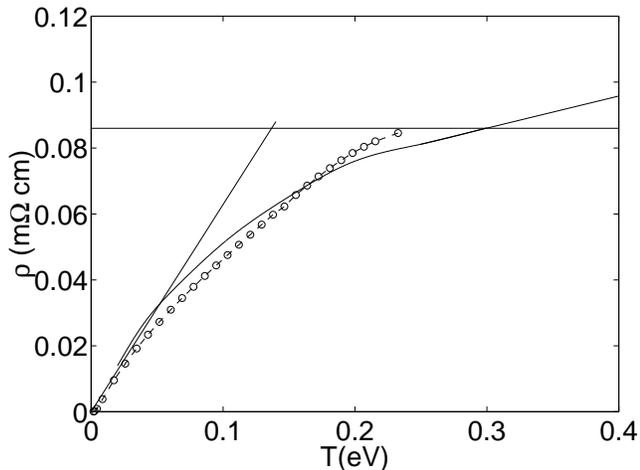}}}}
\caption[]{\label{fig:Nb}Resistivity $\rho(T)$ as a function of
temperature $T$ for Nb according to a semiclassical calculation.
 The figure compares the semiclassical 
(full curve) calculation for $N=640$ with experimental results
(circles).\cite{Nb} It shows the small (Eq. (\ref{eq:d15})) and  
large (Eq. (\ref{eq:f4})) temperature results. The figure 
illustrates that there is saturation also for Nb at large $T$ 
in good agreement with experiment. The large $T$ result ((Eq. 
(\ref{eq:f4})) is slightly exceeded for large $T$, since this 
result is approximate and becomes weakly $T$ dependent in a careful
analysis (Sec. \ref{sec:de}).}
\end{figure}

Fig. \ref{fig:Nb} compares the semiclassical results for Nb with
experimental results. The figure shows a surprisingly good 
agreement between theory and experiment, given the simplicity 
of the model and the absence of adjustable parameters. 
The figure illustrates that saturation also happens for Nb, but at 
a much larger temperature scale than for Nb$_3^{\ast}$. The reason 
for this difference is discussed in Sec. \ref{sec:dd}.

\subsection{C$_{60}$ model}\label{sec:ec}

\begin{figure}
\centerline{
\rotatebox{-90}{\resizebox{!}{3.5in}{\includegraphics{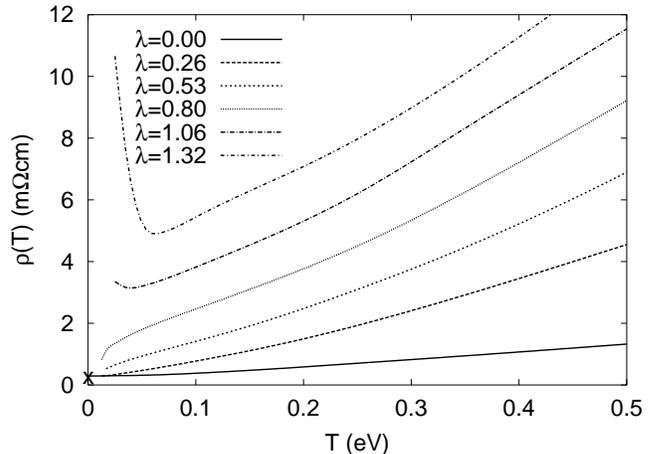}}}}
\caption[]{\label{fig:resistivity01}Resistivity $\rho(T)$ as a function of
temperature $T$ and electron-phonon coupling $\lambda$ for the C$_{60}$ 
model according to QMC calculations. 
The phonon frequency is $\omega_{ph}=0.1$ eV. The figure 
illustrates the lack of saturation. For $\lambda=0.80$ the onset of 
superconductivity can be seen as a sharp downturn in $\rho(T)$ 
as $T$ is lowered, due to superconducting fluctuations. For 
$\lambda=1.06$ and 1.32, the resistivity has a negative slope for small $T$, 
indicating an insulating system. The $x$ shows the resistivity due 
to orientational disorder. }
\end{figure}

Fig. \ref{fig:resistivity01} shows QMC calculations for the 
resistivity of the C$_{60}$ model according to the QMC calculations. 
It illustrates that there is no sign of saturation. Actually 
the curves tend to bend slightly upwards for large $T$. The $x$
indicates the resistivity due to the orientational disorder.
This $T=0$ resistivity was calculated from Eq. (\ref{eq:me3}), i.e.,
independently of the QMC formalism. The curve for $\lambda=0.80$ shows  
signs of superconductivity at small $T$, since the curve turns sharply 
downwards as $T$ is lowered, due to superconducting fluctuations.
For a still larger value of $\lambda$ the system becomes an insulator,
as illustrated by the negative slope of $\rho(T)$ for small $T$.

The solid curve show the result for $\lambda=0$. In this case the 
resistivity is entirely due to the orientational disorder of
the C$_{60}$ molecules. It is interesting that this ``$T$-independent''
scattering mechanism gives rise to a weak $T$-dependence. The reason
for this are discussed in Sec. \ref{sec:de}.

The results for Rb$_3$C$_{60}$ in Fig. \ref{fig:c601} were
measured at a constant pressure and show an approximately quadratic 
dependence on $T$. If these results are converted to a constant
volume measurement, however, an approximately linear dependence on $T$ is
found down to $T \sim 100-200$ K. In agreement with this, Fig.
\ref{fig:resistivity01} $\rho(T)$ shows a rather linear dependence for
$\lambda \le 0.8$ until the superconductivity fluctuations set in. 
The reason for this behavior have been discussed earlier.\cite{Han}

\subsection{Comparison of HI and LE coupling}\label{sec:ed}

The results for the TM and C$_{60}$ models differ drastically.
While the TM model shows saturation, the C$_{60}$ model 
does not. It is interesting to ask to
what extent this is due to a difference in the electron-phonon
coupling (HI versus LE coupling) and to what extent it is due
to other differences, such as the size of the unit cell, the 
lattice structure and 
the band width. For this reason we have also studied the C$_{60}$
model assuming a HI coupling.

The HI coupling in C$_{60}$ is due to intermolecular
phonons, describing the rigid vibrations of the C$_{60}$
molecules relative to each other. The coupling to these phonons
has usually been assumed to be weak.\cite{revmod} This is
also what we find here. We therefore artificially increase the coupling
until $\lambda$ becomes the same as for the intramolecular 
coupling. Since $\lambda\sim \omega_{ph}^{-2}$ for intermolecular 
phonons, we can obtain the increased coupling by artifically reducing 
the phonon frequency $\omega_{ph}$. Experimentally, the intermolecular 
frequencies fall in the range from zero and up to almost 7 
meV.\cite{Pintschovius} We have used a value of $\omega_{ph}=1.8$ meV
which is substantially smaller than the average frequency of the 
experimental spectrum. The resulting $\lambda \sim 0.6$ should 
therefore be substantially larger than the experimental value. 

\begin{figure}
\centerline{
\rotatebox{-90}{\resizebox{!}{3.5in}{\includegraphics{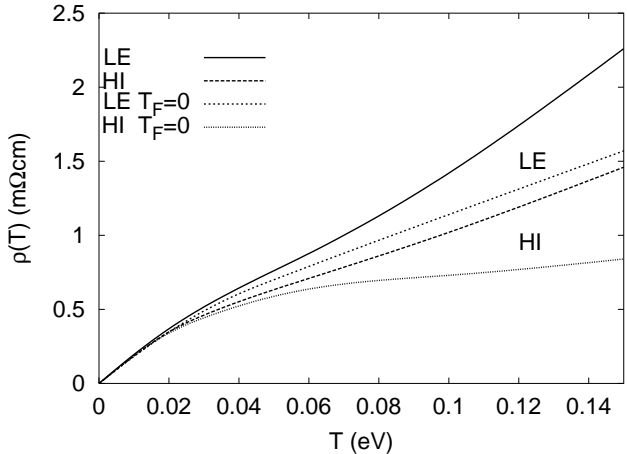}}}}
\caption[]{\label{fig:c60tm}Resistivity $\rho(T)$ as a function of
$T$ for the C$_{60}$ model considering coupling to the level energies
(full line, LE coupl.) and coupling to the hopping integrals
(broken line, HI coupling) according to semiclassical calculations. 
The C$_{60}$ molecules are ordered. The figure also shows results for 
the case when the temperature $T_F$ of the Fermi functions in Eq. 
(\ref{eq:me3}) is put equal to zero. The figure illustrates 
that there is a large difference between LE and HI coupling for 
the C$_{60}$ model.}
\end{figure}

We compare the resistivity in semiclassical calculations for the 
C$_{60}$ model with LE and HI coupling in Fig. \ref{fig:c60tm}. 
The same values of $\lambda\sim 0.6$ and $\omega_{ph}=1.8$ meV 
were used in both cases. The 
molecules are orientationally ordered. While the resistivity
shows now sign of saturation for the LE coupling (full curve), 
the model with HI coupling shows a weak saturation (broken curve). 
This becomes even more pronounced if we neglect the rather trivial 
temperature dependence of the Fermi-functions in Eq. (\ref{eq:me3}).
The resistivity then becomes almost constant for HI coupling and 
large $T$ (dotted curve). For the TM model we find a change of slope 
in $\rho(T)$ for both HI and LE coupling, but the change is more
pronounced for HI coupling.

\section{Discussion}\label{sec:d}

\subsection{Loss of Drude peak}\label{sec:dca}
We mainly focus on temperatures which are so large that the Drude 
peak is essentially lost. The Drude peak is related to intraband
transitions between states with similar ${\bf k}$-vectors. In Appendix 
\ref{sec:da} we illustrate that for Nb$_3$Sb in the semiclassical 
approximation, ${\bf k}$-conservation is lost already at rather 
small values of $T$ and that the concept of intraband transitions 
becomes rather ill-defined. Indeed, for large values of 
$T$, it becomes a good approximation to assume that all states 
couple with the same strength via the current operator to all other
states,\cite{saturationprl} as is illustrated in Appendix \ref{sec:db} 
and in Fig. \ref{fig:dsigma}. The Drude peak is then completely lost. 
Fig.  \ref{fig:dsigma} shows that for Nb$_3$Sb the Drude peak is almost 
completely gone at $T=0.1$ eV.
\begin{figure}[bt]
\centerline{\rotatebox{0}{\resizebox{!}{2.15in}{\includegraphics
{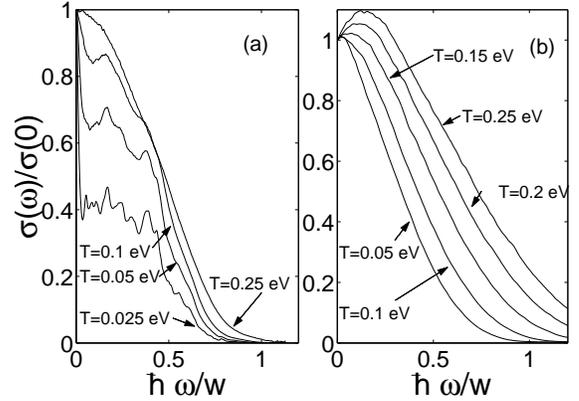}}}}
\caption[]{\label{fig:dsigma}
The optical conductivity as a function of
the frequency $\omega$ for the
(a) A15 and (b) C$_{60}$  models in the semiclassical calculation.
The frequency has been scaled  by the $T=0$ band width $W$.
(a) also shows (broken curve) the result of approximating all current
matrix elements by their average (Eq. (\ref{eq:d2a})).  }
\end{figure}

\subsection{\lowercase{f}-sum rule}\label{sec:dc}
\begin{figure}[bt]
\centerline{\rotatebox{0}{\resizebox{!}{1.6in}{\includegraphics
{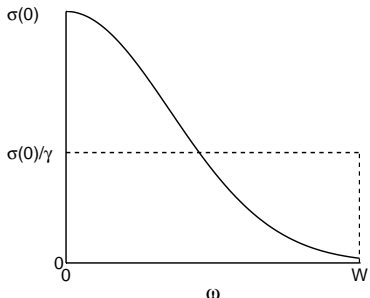}}}}
\caption[]{\label{fig:square}Schematic picture of $\sigma(\omega)$.        
The average over the band width is given by $\sigma(0)/\gamma$.
}
\end{figure}

In the large $T$  limit, the f-sum rule provides a very useful tool for
analyzing the resistivity. For model Hamiltonians of the type
considered here, the f-sum rule takes the form\cite{Maldague}
(for a derivation, see Appendix \ref{sec:f})
\begin{equation}\label{eq:f1}
{2\over \pi}\int_0^{\infty} \sigma(\omega)d \omega=
-{1\over 3}{d^2e^2 \over N\Omega\hbar^2}\langle 0|\hat T_K|0\rangle,
\end{equation}
where $\hat T_K$ is the kinetic energy operator, $d$ is the nearest 
neighbor distance and $\Omega$ is the volume per atom. As discussed 
above, we assume that $T$ is large enough that the Drude peak has 
been smeared out and that $\sigma(\omega)$ is a smooth function. 
We furthermore assume that $\sigma(\omega)=0$ for $\hbar |\omega|>W$,
where $W$ is the band width. This is exactly true in the semiclassical
treatment and approximately true in the QMC treatment. If 
$\sigma(\omega)\equiv \sigma(0)$ for $\hbar|\omega|\le W$, the integral
on the left hand side of Eq. (\ref{eq:f1}) would be
$W\sigma(0)$ and $\sigma(0)$ would simply be given by this integral
divided by $W$. This is shown schematically in Fig. \ref{fig:square}.
For a more general shape of $\sigma(\omega)$ we write
\begin{equation}\label{eq:f2}
\sigma(\omega=0)={\gamma \over W}\int_0^{\infty} \sigma(\omega)\hbar d \omega,   
\end{equation}
where $\gamma$ depends on the shape of $\sigma(\omega)$. To estimate 
$\gamma$ we assume a certain density of states (DOS) $N(\varepsilon)$
and constant matrix elements of the current operator, as discussed
in Appendix \ref{sec:db}. In Table \ref{table:d1} we give the 
value of $\gamma$ for different shapes of $N(\varepsilon)$, namely 
a constant  
\begin{equation}\label{eq:d8}
N(\varepsilon)=\cases {{1\over W}, & if $|\varepsilon|\le W/2$;\cr
0, &  otherwise,\cr}
\end{equation}
a Gaussian
\begin{equation}\label{eq:d9}
N(\varepsilon)={2\over W\sqrt{\pi}}e^{-(2\varepsilon/W)^2}
\end{equation}
and a semi-elliptical 
\begin{equation}\label{eq:d10}
N(\varepsilon)=\cases{8\sqrt{(W/2)^2-\varepsilon^2}/(\pi W^2), 
& if $|\varepsilon|\le W/2$;\cr
0, &  otherwise.\cr}
\end{equation}
DOS. The Table illustrates that $\gamma$ does not depend strongly 
on the shape of the DOS. In the following we assume a semi-elliptical
DOS.

\begin{minipage}{3.375in}                         
\begin{table}
\caption[]{\label{table:d1}The quantity $\gamma$ (Eq. (\ref{eq:f2}))          
and $\alpha$ (Eq. (\ref{eq:f3}))
for a constant (Eq. (\ref{eq:d8})), a Gaussian 
(Eq. (\ref{eq:d9})) and a semi-elliptical (Eq. (\ref{eq:d10}))
density of states (DOS) and for half-filling.}               
\begin{tabular}{cccc}
 & Constant & Gaussian & Semi-elliptical   \\
\hline
$\alpha$ & 0.125 & 0.141 & 0.106 \\
$\gamma$ & 1.44 & 1.81   & 1.91   \\
$\alpha\gamma$& 0.180 & 0.255 & 0.200  
\end{tabular}
\end{table}
\end{minipage}

It is also interesting to study the filling dependence. This is
shown in Table \ref{table:d1a}. The dependence is weak around half-filling,
but $\gamma$ becomes larger for a small filling. 

\begin{minipage}{3.375in}
\begin{table}
\caption[]{\label{table:d1a}The quantities $\gamma$ (Eq. (\ref{eq:f2}))
and $\alpha$ (Eq. (\ref{eq:f3})) for a semi-elliptical DOS 
(Eq. (\ref{eq:d10})) as a function of the fractional filling $p$.
The results are symmetrical around half-filling ($p=0.5$).}
\begin{tabular}{rrrrrr}
$p$  & 0.1      & 0.2      & 0.3    & 0.4  & 0.5     \\
\hline
$\alpha$ & .041 & .070 & .090 & .102 & .106  \\
$\gamma$ & 2.63  & 2.19  & 2.02  & 1.93  & 1.91   \\
$\alpha\gamma$ & .108 & .153 & .182 & .197 & .202 \\
\end{tabular}
\end{table}
\end{minipage}

\subsection{Large $T$ behavior}
As above, we consider temperatures which are so large that the 
Drude peak is gone, but we furthermore assume that the temperatures   
are small compared with the band width. This 
applies, in particular to many transition metal compounds, e.g.,
the A15 compounds.  We consider noninteracting electrons, which should
be a reasonable assumption for broad band transition metal compounds.  
To apply the analysis above, we have to calculate the kinetic energy
$T_K$.  Since $T\ll W$, we can assume $T=0$ in the calculation of
$T_K$. We find that
\begin{equation}\label{eq:f3}
T_K=2n\int_{-W/2}^{\mu}\varepsilon
N(\varepsilon)d\varepsilon\equiv -2n\alpha W N
\end{equation}
is proportional to the band width $W$ and the orbital degeneracy.
The shape of the DOS $N(\varepsilon)$ and the filling enter via
the parameter $\alpha$. This parameter is given in Table \ref{table:d1} 
for different shapes of the DOS for half-filling and in Table 
\ref{table:d1a} for different fillings and a semi-elliptical DOS.
Inserting the result for $T_K$ in the f-sum rule (Eq. (\ref{eq:f1}))
and using Eq. (\ref{eq:f2}), we obtain 
\begin{equation}\label{eq:f4} 
\sigma(0)={\pi \alpha\gamma \over 3} 
{d^3\over \Omega}{ne^2\over \hbar d}.
\end{equation}
Here $\pi \alpha\gamma/3$ depends on the details of the electronic 
structure and is of the order of 0.2, $d^3/\Omega$ depends on the 
lattice structure (see Table \ref{table:d2}), but is of the order 1.

The result (\ref{eq:f4}) is independent of
the band width. This follows, since the kinetic energy 
(\ref{eq:f3}) is proportional to $W$ and is cancelled
by the $W$ in Eq. (\ref{eq:f2}). 

The quantity $ne^2/(\hbar d)$ has the dimension of a conductivity 
and it contains the essential material parameters $n$ and $d$.
For a transition metal compound, with $n=5$ and $d\sim 3$ \AA,
this leads to an upper limit for the resistivity of the order
of 0.1-0.2 m$\Omega$cm. This agrees with the saturation 
resistivities observed for these systems.

\begin{minipage}{3.375in}                         
\begin{table}
\caption[]{\label{table:d2}The quantity $\Omega/d^3$ for different     
lattices, where $d$ is the nearest neighbor distance and $\Omega$
is the volume per atom.
}
\begin{tabular}{ccccc}
   &  fcc  & bcc &   A15 & sc  \\
$\Omega/d^3$ & ${1\over \sqrt{2}}=0.707$ & ${4\over 3\sqrt{3}}=0.770$& 
${4\over 3}=1.333$  & 1   \\ 
\end{tabular}
\end{table}
\end{minipage}

Using the definition of the mean free path $l$ in the introduction 
(Eqs. (\ref{eq:i1})), we can convert the conductivity in Eq. 
(\ref{eq:f4}) to a mean free path 
\begin{equation}\label{eq:d13}
l=cn^{1\over 3}d,
\end{equation}
where for simplicity we have assumed that there is only one 
spherical Fermi surface.  For a semi-elliptical DOS and half-filling 
$c=0.74$ (fcc), 0.72 (bcc) and 0.60 (A15). Thus the quantity
$cn^{1\over 3}$ is close to unity for $n=5$, as appropriate here.
This provides a quantum-mechanical derivation of the Ioffe-Regel 
condition for weakly correlated systems.

In particular for the A15 lattice, the second nearest neighbor
hopping plays  a rather important role. The separation ($0.612a$)
is not much larger than for the nearest neighbors ($0.5a$), but there
are eight second nearest neighbors but just two nearest neighbors.
For this reason, we also define a distance $d$ which is a weighted
average of these distances. As weight factors we use the hopping matrix
elements. Thus we define
\begin{equation}\label{eq:d14}
\langle d^2 \rangle ={\sum_{\nu\mu} d_{\nu\mu}^2 t_{\nu\mu}^2 
\over \sum_{\nu\mu} t_{\nu\mu}^2},
\end{equation}
where $d_{\nu\mu}$ is the distance between the atoms with the 
orbitals $\nu$ and $\mu$. At $T=0$ this increases $d$ from $0.5a$
to about $0.57a$ for the A15 structure. For a semi-elliptical DOS and
filling 0.4, this leads to a larger saturation conductivity and a 
smaller resistivity of about 0.11 m$\Omega$cm instead of 0.14 m$\Omega$cm
if the nearest neighbor separation is used. This is in better 
agreement with the calculated resistivity.

It is interesting to study the filling dependence, indicated
by Table \ref{table:d1a}. We consider Sc, which is the first
element in the $3d$ series. According to a band structure
calculation, Sc has about 1.8 $3d$ electrons.\cite{Papa}
Compared with a system close to half-filling, such as Nb$_3$Sb,
we then expect the saturation resistivity to be about a factor
of 1.5 larger. If we take into account the second nearest neighbor 
hopping, the geometrical factor $<d^2>/\Omega$ is similar for 
the A15 compounds and Sc, suggesting that the filling dependence 
is the dominating factor.
Indeed, while the saturation resistivity
is estimated to be 0.15 m$\Omega$cm for Nb$_3$Sb,\cite{Fisk}
it is well over 0.2 m$\Omega$cm for Sc,\cite{Sc} in agreement
with the expectations. Similar results are also found for
Y.\cite{Y} For the other end members of the $3d$, $4d$ and $5d$ series
clear saturation does not seem to have been observed.

In a similar way we can use the f-sum rule to estimate the resistivity
for the C$_{60}$ model, although the assumption $T\ll W$ is now much more 
questionable, as discussed in Sec. \ref{sec:de}. Considering a fcc 
lattice, using $\gamma=1.91$ and $d=10$ \AA, we obtain
\begin{equation}\label{eq:f4a}
\rho(T)={0.288 \over T_K(T)/(NW)} \hskip0.3cm {\rm m}\Omega{\rm cm}.
\end{equation}
Using the band width $W=0.6$ eV and obtaining $T_K(T)$ from
semiclassical calculations for the C$_{60}$ model, we find       
the saturation resistivity 0.4 m$\Omega$cm.   
The calculated $\lambda=0$ and $T=0$ resistivity (0.29 m$\Omega$cm) 
is below this value, while the results for larger values
of $\lambda$ and $T$ strongly exceed the saturation resistivity. 
The reasons for this are discussed in Sec. \ref{sec:de}.

\subsection{Small $T$ behavior}\label{sec:dd}
In view of the discussion above, we expect the resistivity 
to have an upper limit for models with noninteracting 
electrons scattered by phonons, unless $T$ is very large. 
In many metals, however, the resistivity increases so slowly 
with $T$, that the corresponding conductivity is much large 
than the limit (\ref{eq:f4}) even at the melting temperature. 
The issue of whether or not the resistivity saturates is then 
not raised. It is therefore of interest to study the low $T$ 
behavior of $\rho(T)$. For $T$ larger than some fraction of 
$\omega_{ph}$ we expect\cite{Grimvall}
\begin{equation}\label{eq:d15}
\rho(T)= 8\pi^2{\lambda T k_B\over \hbar \Omega_{pl}^2},
\end{equation}
where $k_B$ is the Boltzmann constant and 
$\lambda$ is the dimensionless electron-phonon coupling constant.
For the TM model with HI coupling we define
$\lambda=\tilde \lambda(\mu,\mu)$, where
\begin{eqnarray}\label{eq:d15a}
&&\tilde \lambda(\varepsilon,\varepsilon^{'}) \\
&&={1\over nKMN(\mu)
\omega_{ph}^2}\sum_{ll^{'}i\alpha}|\langle l|{\partial H\over \partial 
R_{i\alpha}}|l^{'}\rangle|^2\delta(\varepsilon_l-\varepsilon)
\delta(\varepsilon_l^{'}-\varepsilon^{'}-\omega_{ph}), \nonumber
\end{eqnarray}
where $K$ is the number of atoms in a unit cell, the $\alpha$ 
summation is over the three coordinates, $M$ is the 
atomic mass, $|l\rangle$ is an eigenstate of $H$ and  
$i$ labels the atoms in the unit cell. $\Omega_{pl}$ is 
the plasma frequency
\begin{equation}\label{eq:d16}
(\hbar\Omega_{pl})^2={e^2\over 3\pi^2}\sum_n \int_{Bz}d^3k \lbrack {\partial
\varepsilon_{n{\bf k}} \over \partial {\bf k}}\rbrack^2\delta
(\varepsilon_{n{\bf k}}-E_F).
\end{equation}
where $\varepsilon_{n{\bf k}}$ is the energy of a state with
the band index $n$ and the wave vector ${\bf k}$ and $E_F$ is
the Fermi energy. $\Omega_{pl}$ depends on the average Fermi velocity.

The straight lines corresponding to Eq. (\ref{eq:d15}) and
Eq. (\ref{eq:f4}) are shown in Fig. \ref{fig:Nb3}. If these 
lines cross in the experimentally accessible temperature range 
we expect saturation.

It is now interesting to compare our models for Nb and Nb$_3^{\ast}$.
We obtain similar values of $\lambda$ for the two cases, 
$\lambda=1.0$ (Nb$_3^{\ast}$) and $\lambda=0.9$ (Nb).
A larger value of $\lambda=1.7$ for Nb$_3^{\ast}$ was estimated by 
Allen\cite{Allen} while a rather similar value was obtained 
for Nb ($\lambda=1.0$) from {\it ab initio} calculations.\cite{Savrasov} 
We observe that $\lambda\sim 1/\omega_{ph}^2$ depends quite sensitively on 
$\omega_{ph}$. Since we have replaced the whole phonon spectrum 
by three Einstein phonons per atom, obtained as the average of
the phonon spectrum of Nb,\cite{Wolf} one should not expect very
accurate values of $\lambda$ in our calculation. 
For the plasma frequency we obtain $\Omega_{pl}=3.6$ eV (Nb$_3^{\ast}$)
and 8.2 eV (Nb), in rather good agreement with {\it ab initio}
calculations 3.4 eV (Nb$_3^{\ast}$)\cite{Mattheiss} 
and 9.5 eV (Nb).\cite{Savrasov}  

The difference in values of $\Omega_{pl}$ for Nb$_3^{\ast}$ and Nb 
alone then leads to a difference by a factor of five in the slope 
of the line from Eq. (\ref{eq:d15}). 
As a result Nb$_3^{\ast}$ shows a very pronounced saturation already
at small $T$, while Nb only shows sign of saturation at rather 
large $T$. The difference is due to the fact that Nb$_3^{\ast}$
has a large unit cell with many bands and many forbidden crossings.
This leads to quite flat bands and to small electron velocities.
The result is a small plasma frequency (Eq. (\ref{eq:d16})) and a 
steep line from Eq. (\ref{eq:d15}). 

An even more dramatic example is $\alpha$-Mn, which has a unit cell
with 58 atoms.\cite{Mnstruc} One should therefore expect a very 
small plasma frequency and a correspondingly early saturation.
Indeed, it is found that the resistivity saturates at about 
$T=60$ K.\cite{Mnres}

In view of the discussion above, Fig. \ref{fig:dsigma}a
and Eq. (\ref{eq:d15}),
it is tempting to write
\begin{equation}\label{eq:d17}
\sigma(\omega=0,T)={\hbar\Omega_{pl}^2\over 8\pi^2\lambda Tk_B}
+\sigma_{sat},
\end{equation}
where the first term describes the Drude peak (Eq. (\ref{eq:d15}))
and the second term is the conductivity in Eq. (\ref{eq:f4})
at saturation. This formula is correct for small $T$ and for 
$T$ which are so large that the Drude peak is gone but very much
smaller than the band width. Eq. (\ref{eq:d17}) is the ``parallel 
resistor'' formula of Wiesmann {\it et al.}.\cite{Wiesmann} 

\subsection{Very large $T$ behavior}\label{sec:de}    

We have so far discussed temperatures which are so large 
that the Drude peak have been washed out, but which are small 
compared with the band width. We now focus on values of $T$ which 
are large enough that the coupling to the phonons causes a 
substantial change in the band width. Such effects are not very
important for typical transition metal compounds, which have large
band widths. They are, however, of substantial interest for the 
C$_{60}$ model, for which the fluctuations in the level position 
become comparable to the band width at values of $T$ which can be 
reached experimentally.         

At such large values of $T$, there is a rather trivial $T$ dependence 
due to the electron temperature, $T_F$, entering in the Fermi-functions 
of Eq. (\ref{eq:me3}). This can be seen by considering the resistivity
due to static disorder. Although this scattering mechanism is 
$T$-independent, the resistivity is, nevertheless, $T$-dependent. 
Expanding the Fermi functions in Eq. (\ref{eq:me3}) in $1/T$, we 
obtain that $\sigma(0)\sim 1/T$ and $\rho(T)\sim T$ for very large
$T$. A similar dependence also enters for the the electron-phonon 
scattering, and it tends to mask some interesting differences between 
level energy (LE) and hopping integral (HI) couplings. In the following,
we therefore freeze the electron temperature, $T_F=0$, and consider 
the limit of a very large phonon temperature, $T_B$, i.e., we
consider a large $T$ but replace the Fermi function by $\Theta$
functions in Eq. (\ref{eq:me3}).  

The band width entering Eq. (\ref{eq:f2}) can be approximately 
expressed in terms of the second moment of the density of states. 
The same is also approximately true for the kinetic energy. We 
therefore focus on the second moment, 
\begin{equation}\label{eq:me3a}
\langle \varepsilon^2 \rangle = \int_{-\infty}^{\infty} N(\varepsilon)
\varepsilon^2 d\varepsilon,
\end{equation}
which can expressed in terms 
of the Hamiltonian
\begin{equation}\label{eq:de0}
\langle \varepsilon^2\rangle={1\over nN}\sum_{\mu\nu}H_{\mu\nu}^2,
\end{equation}
where $N$ is the number of atoms in the system. 

We first consider the case of the HI coupling. In our semiclassical
formalism we can write
\begin{eqnarray}\label{eq:de1}
&&\sum_{\mu\nu}H_{\mu\nu}(T)-\sum_{\mu\nu}H_{\mu\nu}(T=0)=
\sum_{\mu\nu i\alpha}{\partial H_{\mu\nu}\over \partial R_{i\alpha}}
\delta R_{i\alpha}(T) \nonumber \\
&&+{1\over 2} \sum_{\mu\nu i\alpha}\sum_{j\beta}
{\partial^2 H_{\mu\nu}\over \partial R_{i\alpha}\partial R_{j\beta}}
\delta R_{i\alpha}(T)\delta R_{j\beta}(T)+ ..., 
\end{eqnarray}
where the summation over $i$ extends over all atoms.
Since the displacements $\delta R_{i\alpha}$ are random, we can assume
that
\begin{equation}\label{eq:de2}
\langle \delta R_{i\alpha}(T)\rangle=0 \hskip0.5cm 
\langle \delta R_{i\alpha}(T)\delta R_{j\beta}(T) \rangle=
\delta_{ij}\delta_{\alpha \beta}\langle R^2 \rangle,
\end{equation}
where $\langle R^2 \rangle=k_BT/(M\omega_{ph}^2)$. We then obtain
\begin{eqnarray}\label{eq:de3}
&&\sum_{\mu\nu}H_{\mu\nu}^2(T)-\sum_{\mu\nu}H_{\mu\nu}^2(T=0)=
\langle R^2 \rangle\sum_{i\alpha\mu\nu}({\partial H_{\mu\nu}\over 
\partial R_{i\alpha}})^2  \nonumber \\
&&+ \langle R^2 \rangle \sum_{\mu\nu i\alpha}
H_{\mu\nu}{\partial^2 H_{\mu\nu}\over \partial R_{i\alpha}^2}+ ...
\end{eqnarray}
Explicit calculations for the TM model show that the second term
tend to partially cancel the first term, while for the C$_{60}$
model it adds to the first term. As a crude approximation
we neglect the second term. The first term can be
approximately related to the electron-phonon coupling $\lambda$.
Integrating Eq. (\ref{eq:d15a}) we obtain
\begin{equation}\label{eq:de4}
NW^2\lambda\approx {1\over nMN(\mu)\omega_{ph}^2}
\sum_{i\alpha\mu\nu}({\partial H_{\mu\nu}\over\partial 
R_{i\alpha}})^2,
\end{equation}
where we have assumed that $\tilde \lambda(\varepsilon,
\varepsilon^{'})\equiv \lambda$. Assuming $N(\mu)=1/W$,
we obtain
\begin{equation}\label{eq:de5}
\sum_{\mu\nu}H_{\mu\nu}^2(T)-\sum_{\mu\nu}H_{\mu\nu}^2(T=0)=
nN\lambda W(T=0)k_B T.
\end{equation}
Assuming that $\langle \varepsilon^2\rangle=W^2/12$, as is 
appropriate for a constant density of states, we obtain
\begin{equation}\label{eq:de6}
W(T)=W(0)\sqrt{1+c_{HI}\lambda {k_B T\over W(T=0)}},
\end{equation}
where $c_{HI}=12$.  Comparison with explicit calculations 
for the C$_{60}$ model shows that a more realistic value  
is $c_{HI}\sim 15$.

\begin{figure}
\centerline{
\rotatebox{-90}{\resizebox{!}{3.5in}{\includegraphics{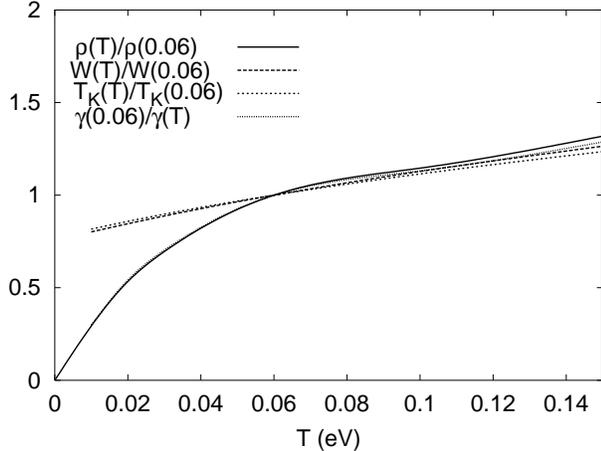}}}}
\caption[]{\label{fig:fa1}
$\rho(T)$, $T_K(T)$, $W(T)$ and $\gamma(T)$ divided by their $T=0.06$ 
eV values for the case of a a coupling to the hopping integrals 
(HI) in the C$_{60}$ model for $\lambda=0.6$ and $\omega_{ph}=0.0018$ 
eV.  The figure illustrates that that the $T$ dependence of $T_K(T)$ and
$W(T)$ are similar and therefore to a substantial extent 
cancel in Eqs. (\ref{eq:f1}, \ref{eq:f2}) for HI coupling in 
the C$_{60}$ model, leading to a weak $T$ dependence of $\rho(T)$ 
for large $T$. The electron temperature $T_F=0$ and only the 
boson temperature $T_B$ is varied.}
\end{figure}

We next consider the kinetic energy, $T_K$. As discussed
in Sec. \ref{sec:dc} (Eq. (\ref{eq:f3})), the kinetic energy is
closely related to the band width via the quantity $\alpha$.
As $T$ is increased, however, the shape of $N(\varepsilon)$ 
changes somewhat, and there is not a perfect proportionality
between $W(T)$ and $T_K(T)$. This is illustrated in Fig. 
\ref{fig:fa1}, where the curves describing the $T$ dependence
of these two quantities differ slightly. Nevertheless, 
from Eqs. (\ref{eq:f1}, \ref{eq:f2}),
it follows that the $T$ dependence of these two quantities
largely cancel in the calculation of $\sigma(\omega=0)$
and $\rho(T)$. This is illustrated in Fig. \ref{fig:fa1},
where $\rho(T)$ has only a weak $T$ dependence, once the 
resistivity has ``saturated'' (at about $T=0.06$ eV). 
The remaining $T$ dependence is due to the $T$ dependence of 
$\alpha$ and $\gamma$.

We next consider the case of the LE coupling for the C$_{60}$
model. In this case the second moment is the sum of one contribution 
from the hopping (off-site elements) of $H_{\mu\nu}$ in Eq. 
(\ref{eq:de0}) and one contribution from the fluctuations of
the level energies (on-site  terms) in Eq. (\ref{eq:de0}). To   
obtain the fluctuations in the on-site terms, we rewrite the 
interaction terms as
\begin{equation}\label{eq:red6a}
H^{\rm el-phon}=\sum_{imm^{'}\sigma}v_{imm^{'}}
\psi^{\dagger}_{im\sigma} \psi_{im^{'}\sigma},
\end{equation}
where $v_{imm^{'}}$ is a boson operator. The average of the
fluctuation in the level position can then be written as
\begin{equation}\label{eq:red6b}
\langle \sum_{mm^{'}\sigma}v_{imm^{'}}^2 \rangle^{1/2}\approx 3\sqrt{
{\lambda\over N(\mu)}T}
\end{equation}
in the limit of a large $T$. Combining this with the off-site 
term gives
\begin{equation}\label{eq:de7}
n\langle \varepsilon^2\rangle_T=n\langle \varepsilon^2\rangle_{T=0}
+{3\lambda \over N(0)}T,
\end{equation}
where $n=3$. Assuming a constant $N(\varepsilon)$, we estimate
that $T_K(T=0)/N\approx -2.6\langle \varepsilon^2\rangle_{T=0}^{1/2}$. 
For a large $T$, the coupling to the phonons leads to large 
separations of the levels, and we can use perturbation theory 
for calculating the kinetic energy.
\begin{equation}\label{eq:de8}
T_K(T)=2\sum_{\mu}^{\rm occ}\sum_{\nu}^{\rm unocc}
{t_{\mu\nu}^2\over \varepsilon_{\mu}-\varepsilon_{\nu}} 
\approx {1\over 2}  {Nn\langle \varepsilon^2\rangle_{T=0}
\over \langle \varepsilon_{\mu}-\varepsilon_{\nu} \rangle },
\end{equation}
where we have replaced the denominator by an average denominator
$\langle \varepsilon_{\mu}-\varepsilon_{\nu} \rangle$ and
the limitations on the sums to occupied and unoccupied states 
introduce a factor of 1/4. A simple estimate of 
$\langle \varepsilon_{\mu}-\varepsilon_{\nu} \rangle$ is obtained
by assuming that the levels have the energies $\pm \Delta \varepsilon/2 $.
Then the separation of the levels is $\Delta \varepsilon=
2  \langle \varepsilon^2\rangle_T^{1/2}$, where only the on-site 
contribution to $\langle \varepsilon^2\rangle$ should be included. 
At large $T$, however, the on-site contribution dominates and we 
have dropped this restriction.  Then
\begin{equation}\label{eq:de9}
T_K(T)\approx -2.6 N {\langle \varepsilon^2\rangle_{T=0}  
\over \langle \varepsilon^2\rangle_T^{1/2}},
\end{equation}
where we have used the same prefactor 2.6 as below Eq. (\ref{eq:de7}).
This gives a better agreement with the numerical results than the
prefactor (3/4) derived from the arguments above,
which is substantially too small, as one would expect. The averaging
in Eq. (\ref{eq:de8}) greatly favors small values of the denominator,
while our simple estimate focuses on large values. The estimate
in Eq. (\ref{eq:de9}) is also a good estimate for $T=0$, as
shown above, and actually for the whole temperature range. As 
usual, we relate the band width to the second moment. Assuming 
a constant DOS, Eqs. (\ref{eq:f1}, \ref{eq:f2}) give
\begin{equation}\label{eq:de10}
\sigma(0)={2.6 \pi \gamma \over 6 \sqrt{12}} {e^2d^2\over \Omega\hbar}
{\langle \varepsilon^2\rangle_{T=0} \over \langle \varepsilon^2\rangle_T},
\end{equation}
where one factor $\langle \varepsilon^2\rangle_T^{1/2}$ comes
from the band width and one factor from the kinetic energy.
Since $\langle \varepsilon^2\rangle_T$ grows with $T$ 
(Eq. (\ref{eq:de7})), both the kinetic energy and the band width 
work together to reduce $\sigma(0)$ and to increase $\rho(T)$ as 
$T$ is increased. Thus we obtain       
\begin{equation}\label{eq:de11}
\rho(T)={0.8\over \gamma(T)}(1+c_{LE} \lambda {k_BT\over W}) 
\hskip0.5cm {\rm m}\Omega{\rm cm}.
\end{equation}

\begin{figure}
\centerline{
\rotatebox{-90}{\resizebox{!}{3.5in}{\includegraphics{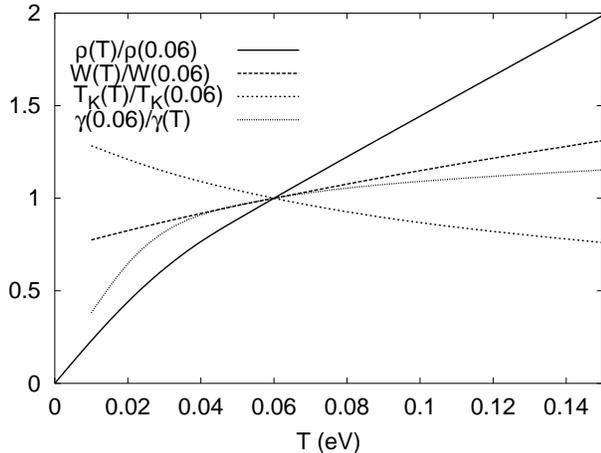}}}}
\caption[]{\label{fig:fa2}
$\rho(T)$, $T_K(T)$, $W(T)$ and $\gamma(T)$ divided by their $T=0.06$ 
eV values for the case of a a coupling to the level energies  (LE) 
in the C$_{60}$ model with $\lambda=0.6$ and $\omega_{ph}=0.0018$ eV.
The figure illustrates that that the $T$ dependence of $T_K(T)$ and
$W(T)$ are the opposite and therefore work in the same direction 
in Eqs. (\ref{eq:f1}, \ref{eq:f2}) for the LE coupling in the 
C$_{60}$ model, leading to a strong $T$ dependence of $\rho(T)$. 
The electron temperature $T_F=0$ and only the boson temperature $T_B$ is 
varied.}
\end{figure}

From the derivation we obtain $c_{LE}=12$. A better fit to the data
is obtained from $c_{LE}=16$. In addition we observe that there
is also an appreciable $T$ dependence in $\gamma(T)$.
These results are illustrated in Fig. \ref{fig:fa2}.
In particular, we notice that $W(T)$ and $T_K(T)$ have the
opposite $T$ dependence, and therefore work together in
the expressions in in Eqs. (\ref{eq:f1}, \ref{eq:f2}).
This is in strong contrast to the case of HI coupling,
where the two $T$ dependencies largely cancel each other.

\subsection{Lack of saturation in the C$_{60}$ model}\label{sec:df}

By using the f-sum rule, we showed in Sec. \ref{sec:dc} that 
one should expect the resistivity of the alkali-doped fullerenes
to saturate at about 0.4 m$\Omega$cm. Actually, this value      
is almost reach already at $T=0$  (0.3 m$\Omega$)
due to the orientational disorder. One can therefore consider
the C$_{60}$ model as a case where saturation has already
happened at $T=0$. 

\begin{figure}
\centerline{
\rotatebox{-90}{\resizebox{!}{4.0in}{\includegraphics{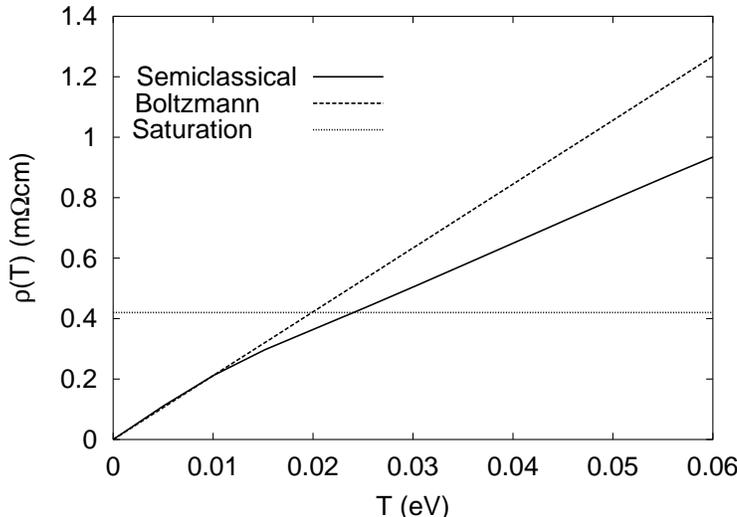}}}}
\caption[]{\label{fig:df1}Resistivity of the ordered C$_{60}$ model
for $\omega_{ph}=0.00001$ eV,    $\lambda=0.6$ and LE coupling.
The figure illustrates that the semiclassical (full curve) and the
Boltzmann (dashed curve) agree well for small $T$, but deviates
when $\rho(T)$ becomes comparable to the hypothetical saturation 
resistivity (horizontal dotted curve). The figure shows that there 
is no real saturation in this case.}
\end{figure}

This can be further illustrated by considering
the resistivity for a model where all the C$_{60}$  molecules
have the same orientation, i.e., a system without disorder.
The results are compared with the resistivity expected from
the Boltzmann equation in Fig. \ref{fig:df1}. The phonon frequency
has been chosen very small, so that the Boltzmann equation gives
a linear behavior for all $T$ of interest. For small values of $T$
the Boltzmann equation and the semiclassical theory agree. However, 
when $\rho(T)$ becomes of the order of 0.3 m$\Omega$cm, shortly 
before saturation might have been expected, the two curves start
to deviate. At this point we may consider the system has having 
saturated, and the theory in Sec. \ref{sec:de} of very large $T$
applies. This theory also predicts a linear behavior, but not 
necessarily with the same slope as at small $T$. Simple arguments suggest 
that the two slopes might be of the same order of magnitude,
as found in Fig. \ref{fig:df1}. The small $T$ slope is, however, 
related to the properties around the Fermi energy, while the
very large $T$ slope refers to properties integrated over all 
states. The two slopes should therefore not be expected to be 
the same. The Boltzmann equation is not qualitatively wrong
for large $T$ in this case, but the relatively good agreement 
for large $T$ is somewhat accidental.

For the disordered C$_{60}$ model, the disorder itself leads 
to a resistivity comparable to the ```saturation resistivity´´,
and the ``very large $T$'' limit in 
Sec. \ref{sec:de} applies already for any finite $T$. This theory
predicts that $\rho(T)$ has a linear dependence on $T$, as is also
approximately seen (see Fig. \ref{fig:resistivity01}). The resistivity
could be considered to have ``saturated'', but this concept is 
meaningless for the C$_{60}$ model, since the resistivity grows
linearly, with a large slope, also after ``saturation''. 

We observe that the boson character of the phonons is important
for the arguments in this section and in Sec. 
\ref{sec:de}.\cite{Han}  Because of this, the number of phonons
grow without limit as $T$ is increased, leading to the
corresponding growth in the phonon amplitude $\langle R^2\rangle$.
This leads to a continuing growth of the band width and reduction
of the kinetic energy for the case of LE coupling. As a result
the resistivity does not saturate. 

This is different from the
case of electron-electron scattering, where Fermi occupation
numbers enter the theory. As a result, we have found that
there is saturation of the resistivity in a simple one-band,
symmetric, half-filled Hubbard model, at least in the dynamical
mean-field theory.\cite{Han} In view of this, it is interesting 
that the High $T_c$ cuprates are usually considered as examples
of systems where the resistivity does not saturate, although
electron-electron scattering is often believed to be the 
dominating mechanism. This issue is addressed in the next section.

\subsection{Saturation for High $T_{\lowercase {c}}$ cuprates}\label{sec:dg}

The resistivity in some of the High $T_c$ cuprates is substantially
larger than one would expect from the Ioffe-Regel 
criterion.\cite{saturationhightc} It has therefore been assumed 
that these compounds are examples of systems where the resistivity
does not saturate. Using the f-sum rule, however, we have found
that the resistivity saturation is to be expected at much higher
resistivities than predicted by the Ioffe-Regel criterion or
what is found for, e.g., the A15 compounds.\cite{saturationhightc}
The reason is that the kinetic energy is strongly reduced in these 
systems. This is partly due to the strong Coulomb interaction
reducing hopping, in particular for systems with a small doping
$x$. Furthermore, only the $x^2-y^2$ orbital is believed to play 
an essential role, leading to a small degeneracy $n=1$. As a result, 
for La$_{2-x}$Sr$_x$CuO$_4$ we find\cite{saturationhightc}
\begin{equation}\label{eq:dg1}
\rho(T)={0.4\over x(1-x)} \hskip0.3cm {\rm m}\Omega{\rm cm}.
\end{equation}
This result is much larger than the saturation resistivity of the 
order of 0.1 m$\Omega$cm for the A15 compounds, in particular for
small $x$. Experimental resistivities are smaller than Eq. 
(\ref{eq:dg1}), but for small values of $x$ not much smaller.\cite{Takagi} 
For these cases signs of saturation are indeed seen.\cite{Takagi}
We therefore conclude that the data are consistent with saturation.
Actually, the data show signs of saturation when the experimental
resistivity comes close to the expected saturation resistivity
(\ref{eq:dg1}).

\subsection{Relation to Mott's minimum conductivity}\label{sec:fb}

Within the semiclassical theory, the phonons cause a static disorder.
The problem discussed here therefore has some relations to the
conduction in disordered system. Thus the LE and HI couplings 
correspond to diagonal
and off-diagonal disorder, respectively. While the disordered systems
are usually studied for small $T$, we are here interested in the large
$T$ behavior. In the semiclassical theory, however, apart from causing 
disorder, $T$ only enters
via the Fermi-functions, and it does not play an important role for 
the qualitative behavior. Below we therefore compare our work
with the treatment of disordered systems.

Diagonal disorder can lead to an Anderson metal-insulator
transition at $T=0$.\cite{Ramakrishnan} For the case of 
off-diagonal disorder, however, Antoniou and Economou\cite{offdiagonal}
have found that there is no metal insulator transition if
the Fermi energy is located in some finite region around
the middle of the band. Our semiclassical calculations
agree with these results, i.e., we find localization for
LE but not for HI coupling as $T$ is increased.

In the QMC calculation of the resistivity, however, we
see no sign of localization for LE coupling, just a lack of 
saturation.  This is natural. Localization depends sensitively
on the phase factors, which are not destroyed in the elastic scattering
in an disordered system. In the inelastic scattering by phonons at
finite $T$ these phase factors are, however, lost, and localization
is not expected.\cite{Ramakrishnan} The effects of the inelastic 
scattering is properly included in the QMC but neglected in the 
semiclassical treatment, and therfore localization shows up in the 
semiclassical but not in the QMC treatment.  

Mott \cite{Mott} has argued that as the disorder increases, 
there is a discontinuous transition from a metal to an insulator 
at $T=0$.  He therefore introduced the concept of the minimum conductivity
\begin{equation}\label{eq:mott1}
\sigma_{\rm min}=0.03 {e^2\over \hbar d},
\end{equation}
where $d$ is the nearest neighbor atomic distance.                    
Later work has argued that the transition from a metal to an insulator
actually is continuous, but that $\sigma_{\rm min}$ may still have 
some relevance for low but nonzero temperatures.\cite{Ramakrishnan}
We therefore make a comparison of $\sigma_{\rm min}$  to the 
resistivity in the TM and C$_{60}$ models. Converting Eq. (\ref{eq:mott1})
to a resistivity, we obtain
\begin{equation}\label{eq:mott2}
\rho_{\rm max}=1.6 d \ \ {\rm m}\Omega{\rm cm},
\end{equation}
where $d$ is measured in \AA.
Based on experiment, Mott deduced a somewhat larger minimum conductivity
for systems containing transition metal atoms, 
resulting in the maximum resistivity
\begin{equation}\label{eq:mott3}
\rho_{\rm max}=1 \ \ {\rm m}\Omega{\rm cm}.
\end{equation}
   
Mott derived his result for diagonal disorder. His result can most
naturally be compared with our saturation resistivity for HI coupling
(off-diagonal disorder), since saturation is most pronounced in this 
case. The resistivity $\rho_{\rm max}$ is much larger than the saturation 
resistivity obtained above (Eq. (\ref{eq:f4})) for the TM model with 
a five-fold degenerate orbital ($n=5$). For a fcc lattice and a 
half-filled semi-elliptical band it takes the form
\begin{equation}\label{eq:mott4}
\rho_{\rm sat}={0.14 d \over n} \ \ {\rm m}\Omega{\rm cm},
\end{equation}
which is of the order of 0.1 m$\Omega$cm. The corresponding 
conductivity is substantially larger than Mott's minimum conductivity.

\subsection{Alternative explanations}\label{sec:fc}

Cote and Meisel\cite{Cote} proposed an interesting explanation of
saturation. They argued that  the electrons would not see phonons 
with a wave length $\Lambda$ that is much longer than the mean free 
path. They therefore assumed that an electron can only be scattered by
a phonon if $l>\Lambda$. As $T$ is increased and $l$ is reduced, an 
increasing fraction of the phonons become inefficient as scattering 
sources. The result is that $\rho(T)$ increases much slower than $T$ at
large $T$, in rather good agreement with experiment.\cite{Cote}
We are now in the position to test this assumption.

Above, we have studied a model with three local Einstein 
phonons on each atom, describing the vibrations in the three
coordinate directions. This is equivalent to study Einstein
phonons in ${\bf q}$-space. We then write the displacement 
of the atom at the unperturbed position ${\bf R}_i^0$ as
\begin{equation}\label{eq:a1} 
\delta {\bf R}_l={1\over \sqrt{N}}\sum_{j \alpha}
{\bf u}_{j\alpha}e^{i{\bf q}_j \cdot {\bf R}_l^0}, 
\end{equation}
where $j=1, ..., N$ labels the $N$ ${\bf q}$-vectors and $\alpha$
labels the three modes for each ${\bf q}$-vector. The corresponding
phonon amplitude is ${\bf u}_{j\alpha}$. We perform a 
calculation where the phonons are treated semiclassically as before, 
but where the amplitudes ${\bf u}_{j\alpha}$ are treated as 
random variables. This gives the same resistivity as before.
We then gradually turn off the long wave length phonons, putting 
the corresponding amplitudes ${\bf u}_{j\alpha}=0$.
For small $T$ we expect this to reduce the resistivity. 
For large $T$, however, the arguments of Cote and Meisel\cite{Cote}
suggests that this should not influence the resistivity if $\Lambda>l$
for the phonons turned off. 

\begin{figure}
\centerline{
\rotatebox{0}{\resizebox{!}{2.5in}{\includegraphics{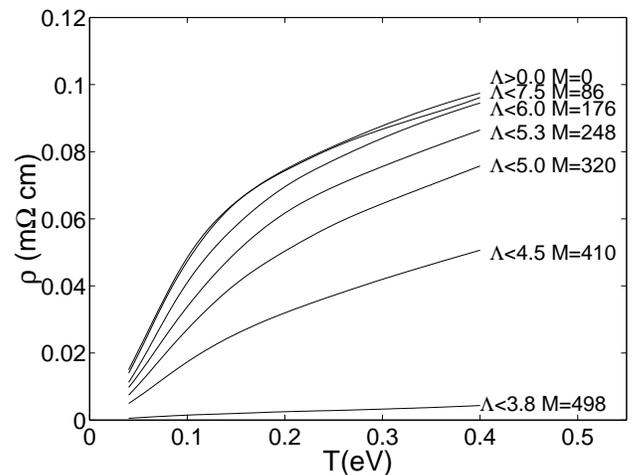}}}}
\caption[]{\label{fig:cote}The resistivity of Nb as a function of $T$.
The scattering from phonons of successively shorter and shorter wave 
lengths $\Lambda$ is suppressed. For the uppermost curve all phonons
are considered. In the lower curves the phonons corresponding 
to the  $M$ shortest ${\bf q}$-vectors (longest
wave lengths) were suppressed, where $M$ is marked at the curve.
The figure illustrates that the long wave length phonons contribute
about equally much both to the small $T$ resistivity and large $T$ 
resistivity.} 
\end{figure}

We group the ${\bf q}$-vectors with equal length in shells. Shells
with ${\bf q}$-vectors of similar length are further grouped together
in such way that each group contains a similar number of ${\bf q}$-vectors.
Then the groups of phonons are successively turned off.
The results are shown in Fig. \ref{fig:cote}. The figure illustrates
that as a group of phonons is turned off there is a drop in the resistivity.
This is not only true for small $T$ but for all $T$ studied here. 
Consider for instance the curve with all phonons included and $T=0.4$ eV.
The resistivity $\rho\sim 0.1$ m$\Omega$cm corresponds to $l\sim 3.5$ \AA.
The theory of Cote and Meisel then assumes that all phonons with
$\Lambda>3.5$ \AA \ can be turned off without $\rho$ changing.
The figure illustrates that this is far from the result of our calculation.
This illustrates that also phonons with a relatively long wave length
contribute substantially to the large $T$ resistivity, although $\Lambda>l$.

Fig. \ref{fig:cote} illustrates that phonons with a very long wave length
make a small contribution to the resistivity for any $T$.
The reason is that a long wave length phonon does not change 
the relative separation of two neighboring atoms very much, which means 
that the corresponding hopping matrix element is not changed very much.

\begin{figure}
\centerline{
\rotatebox{-90}{\resizebox{!}{3.5in}{\includegraphics{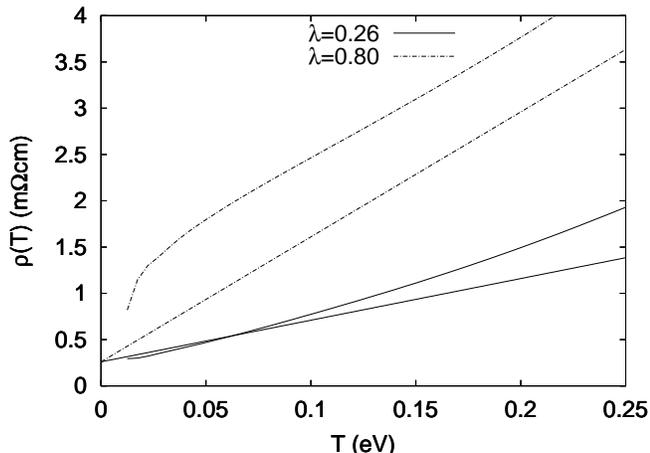}}}}
\caption[]{\label{fig:linearity}Resistivity $\rho(T)$ as a function of
temperature $T$ and electron-phonon coupling $\lambda$ for the C$_{60}$ 
model according to QMC calculations. The phonon frequency is 
$\omega_{ph}=0.1$ eV. The straight lines 
show the resistivity $\rho(T)=0.29+17\lambda T$ m$\Omega$cm, where 
$0.29$ is the resistivity due to the orientational disorder. The figure 
illustrates that there is some ``excess'' resistivity at moderate $T$
and large $\lambda$ but no saturation for this model.}
\end{figure}

It has also been argued\cite{Millis} that resistivity saturation can be 
understood in a Holstein model, somewhat similar to our C$_{60}$
model. For small $T$ and large $\lambda$ the Holstein model shows 
an ``excess'' resistivity. Similar 
effects are observed in our C$_{60}$ model, as is seen in Fig. 
\ref{fig:resistivity01} for $\lambda=0.80$. The result is that 
the slope of the $\rho(T)$ curve is reduced as $T$ is increased. 
To analyze this, we compare the calculated $\rho(T)$ with the 
resistivity
\begin{equation}\label{eq:r1}
\rho(T)=0.29+17  \lambda T \hskip0.3cm {\rm m} \Omega {\rm cm},
\end{equation}
in Fig. \ref{fig:linearity}. The value 0.29 comes from the orientational
disorder and the term $\sim \lambda T$ is the type of behavior we expect
for a normal nonsaturating system (e.g., from Boltzmann theory). 
The slope was adjusted to the results for $\lambda=0.26$.
For such a small value of $\lambda$ there is no sign of saturation 
in Fig. \ref{fig:resistivity01}. If the system shows saturation
for larger values of $\lambda$, we would then expect the calculated
resistivity to be below Eq. (\ref{eq:r1}). We find, however,
that QMC results for large $T$ stay above these results  
for all values of  $\lambda$ that we have studied. In
the figure this is illustrated for $\lambda=0.8$. As pointed out in
Ref. \onlinecite{Millis}, the resistivity in this model actually 
does not   saturate, and it was concluded that ``saturation'' is 
a misnomer. As we have shown above, however, the TM model 
is a much better model of saturation, both because it is much more 
realistic for systems showing saturation, and because it also gives 
results much more similar to experiment.

In a semiclassical treatment of the type used by Millis 
{\it et al.} the ``excess'' resistivity for large $\lambda$ and 
small $T$ is due to the formation of a highly anharmonic potential 
well for the phonons. This leads to a larger vibration amplitude 
and an increased resistivity. Similar results are found in our 
QMC calculation, as discussed above. In a more realistic model, 
the electrons would couple to many phonon modes, each typically with 
a substantially weaker coupling. Even if the total $\lambda$ may be 
large, each  phonon would in such a model have a more harmonic potential
well, and we would not expect a large ``excess'' resistivity.
This further supports our belief that this type of model is not
appropriate for describing resistivity saturation.

\section{Summary}
We have studied  models of weakly correlated transition metal 
compound (TM model) and of alkali-doped fullerenes (C$_{60}$ 
model). These models were studied using Quantum Monte-Carlo (QMC) 
and semiclassical methods. The results, as well as earlier results
for the High $T_c$ cuprates, were analyzed by using the f-sum rule.
We assumed that $T$ is so large that the Drude peak has been smeared 
out. Then (Eqs. (\ref{eq:f1},\ref{eq:f2})) an approximate lower
limit to $\sigma(0)$ is given by
\begin{equation}\label{eq:s1}
{1\over \rho(T)}=\sigma(0)\sim {1\over W}\int_0^{\infty}\sigma(\omega)
d\omega\sim {|T_K(T)| \over d W(T)},
\end{equation}
where $T_K(T)$ is the kinetic energy, $W(T)$ is the band width 
and $d$ is the nearest neighbor distance. 

We first considered $T\ll W$. For the TM model of noninteracting 
electrons, it then followed that $T_K\sim W$. This leads to the 
simple upper limit 
\begin{equation}\label{eq:s2}
\sim {\hbar d\over n e^2}
\end{equation}
for the resistivity, where $n=5$ is the orbital degeneracy of the
$d$-level. This agrees rather well with the saturation resistivity of many
transition metal compounds, and it corresponds to a mean free 
path $l\sim d$.

For the High $T_c$ compounds, the kinetic energy is strongly 
reduced by correlation effects. There is a strong reduction
in the hopping probability of a hole to a neighboring site
if there already is a hole on this site. This leads to
$|T_K|\sim x(1-x)$, where $x$ is the doping. The corresponding
upper limit for the resistivity is then 
\begin{equation}\label{eq:s3}
\sim {\hbar c\over e^2 x(1-x)}
\end{equation}
where $c$ is the distance between two CuO$_2$ planes. 
Since essentially only the $x^2-y^2$ orbital is involved, 
the degeneracy factor is $n=1$. This resistivity is 
much larger than for the TM model, both because of $n=1$ and 
because of factor $x(1-x)$. This limit is therefore apparently 
never exceeded for any high-$T_c$ compound. There are only a 
few cases where the resistivity gets close to this limit, and
in these cases the resistivity shows signs of saturation.

Whether or not saturation is actually observed, depends on how rapidly 
the resistivity grows for small $T$'s. In this limit we have
$\rho(T)\sim \lambda T/\Omega_{pl}^2$ for the TM model. 
For the A15 compounds, e.g., Nb$_3$Sn, $\lambda$ is fairly large and 
$\Omega_{pl}$ is very small, due to the large unit cell and the quite 
flat bands. The result is that the resistivity
grows very rapidly for small $T$ and gets close to the limiting value
for rather small $T$. The resistivity then shows a pronounced
saturation. For Nb, on the other
hand, $\Omega_{pl}$ is much larger and the resistivity grows much more
slowly with $T$, and there is only a weak saturation. For most 
metals, the limiting resistivity would only be reached far above
the melting temperature, due to the slow increase of $\rho(T)$
for small $T$. 

We also considered very large values of $T$, where $T$ becomes 
comparable to the band width. Then both $T_K$ and $W$ have 
strong $T$ dependences. It is important to distinguish
between the case when the phonons couple to the level
positions (LE coupling) and to the hopping integrals (HI coupling).
In the former case, $T_K$ decreases with $T$, since the different
levels have different energies, and hopping is reduced. In the
latter case, $T_K$ is increased, since the square of the hopping
integrals increases with $T$. In both cases $W$ increases with $T$. 
In the LE case, both effects work together (Eq. (\ref{eq:s1}))
to reduce $\sigma(0)$ and to increase the resistivity. In the HI case,
on the other hand, the two effects partly compensate each other,
and the increase in the resistivity is smaller. 

These considerations are very relevant for the C$_{60}$ case.
Due to the orientational disorder, the saturation limit can
be considered to have been reached already for $T=0$.
Because of the the small band width, however, the $T$ dependence
of the band width and the kinetic energy become very important.
Furthermore, the coupling is of the LE type, so that the 
$T$ dependence of these two quantities cooperate in increasing
the resistivity. The result is a drastic increase in the 
resistivity, beyond the ``saturation resistivity'', 
and little or no sign of saturation.
We may therefore consider C$_{60}$ to belong to a different 
class than the A15 and High $T_c$ compounds.

\begin{figure}
\centerline{
\rotatebox{-90}{\resizebox{!}{3.5in}{\includegraphics{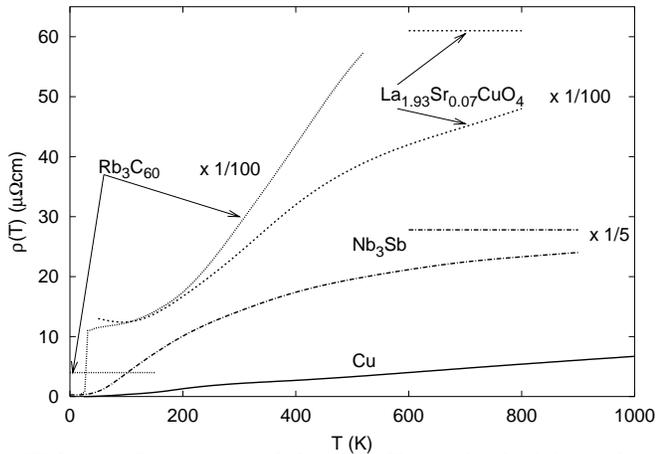}}}}
\caption[]{\label{fig:all}Resistivity of Cu, Nb$_3$Sb (multiplied by
a factor 1/5),\cite{Fisk} La$_{1.93}$Sr$_{0.07}$CuO$_4$ (multiplied 
by 1/100),\cite{Takagi} and alkali-doped C$_{60}$ 
(multiplied by 1/100).\cite{Hebard} The figure also shows our estimated 
saturation resistivities for the latter three cases, with the C$_{60}$
saturation resistivity ($\sim$ 0.4 m$\Omega$cm) barely visible at about 
$\sim$400/100=4   $\mu\Omega$cm. The figure illustrates that the 
resistivity saturates for Nb$_3$Sb and La$_{1.93}$Sr$_{0.07}$CuO$_4$ 
but not for alkali-doped C$_{60}$.} 
\end{figure}
This is illustrated in Fig. \ref{fig:all}, which shows the resistivity 
for Cu, Nb$_3$Sb,  La$_{1.93}$Sr$_{0.07}$CuO$_4$ and hole-doped
C$_{60}$, where the resistivities of the latter three metals have
been reduced by factors 5, 100 and 100. The resistivities of 
Nb$_3$Sb and  La$_{1.93}$Sr$_{0.07}$CuO$_4$ stay below the 
expected saturation resistivities, while the resistivity of 
C$_{60}$ is far above the ``saturation'' resistivity, shown in the
lower left corner of the figure. This suggests that the systems studied
here fall in three different classes, namely i) weakly correlated 
transition metal compounds, showing saturation in agreement with the
Ioffe-Regel condition, ii) strongly correlated high-$T_c$ cuprates, 
showing saturation but at much larger values than predicted by 
the Ioffe-Regel condition, and iii) alkali-doped C$_{60}$ compounds, 
showing no saturation.

We would like to thank M. F\"ahnle, O. Jepsen, P. Horsch, B. Keimer
and R. Zeyher  for useful discussions, M. Jarrell for making his 
maximum entropy program available and the Max-Planck-Forschungspreis for 
financial support.

\appendix

\section{QMC for the TM model}\label{sec:TM}
The Hamiltonian of the TM model can be written as 
\begin{equation}
H=\sum_{\mu,\nu} H_{\mu,\nu}=\sum_{\eta=1}^{N_b} H_{\eta}
\end{equation}
where $\eta$ labels a given ordering of the $N_b = z N n^2 $ 
bonds, $z$ being the number of atoms connected to a given site 
by the Hamiltonian operator.

Defining $\Delta \tau = \beta/L$, using Trotter decomposition at lowest
order and breaking up the Hamiltonian in $N_b$ terms, 
the partition function is, \cite{footbreakup}
\begin{equation}
Z= Tr \left[\prod_{l=1}^{L} e^{-\Delta \tau H} \right] \simeq Tr \left[\prod_{l=1}^{L} \prod_{\eta=N_b}^{1} e^{-\Delta \tau H_{\eta}}\right] \label{eq:breakup}
\end{equation}
Integrating out the electron degrees of freedom \cite{Scalapino} leads to
\begin{equation}
Z=\left[\det(1+B_L B_{L-1}...B_1)\right]^2
\end{equation}
with
\begin{equation}
B_l= \prod_{\eta=N_b}^{1} b_{\eta}^{l} = \prod_{\eta=N_b}^{1} e^{-\Delta \tau H_{\eta}}
\end{equation}
The matrices $b^{l}_{\eta}$ have dimension $N n$ and have the following
form:
\begin{equation}
b^{l}_{\eta}=\left(\begin{array}{ccccccc}
1 & .. & 0 & .. & 0 & .. & 0 \\
 :&:&:&:&:&:&:\\
0 & .. & \cosh(\Delta \tau H_{\eta})& .. & \sinh(-\Delta \tau H_{\eta})& ..&0\\
 :&:&:&:&:&:&:\\
0 & .. & \sinh(-\Delta \tau H_{\eta})& .. & \cosh(\Delta \tau H_{\eta})& ..&0\\
 :&:&:&:&:&:&:\\
0 & .. & 0 & .. & 0 & .. & 1
\end{array}\right)\label{eq:smallb}
\end{equation}
It can be shown \cite{Scalapino} that the electron Green function is 
written as:
\begin{equation}
g=(1+B_L...B_2 B_1)^{-1} \label{eq:green}
\end{equation}
During the simulation, $g$ and $g^{-1}$ are constantly stored 
and updated.

A Quantum Monte Carlo move is a displacement of a phonon coordinate for a
given slice. The move is then accepted or rejected according to the
Metropolis algorithm which involves the calculation of the square 
determinant ratio between the electron Green functions after and before 
the displacement, $R^2=\left[\det(g')/\det(g)\right]^2$.

Without loss of generality let us suppose that an atom $i$ is displaced 
in the first slice (so that we can omit the higher index in $b^{1}_{\eta}$). 
This will involve a change in $B_1 \to B'_1$ or $N_c=z n^2$ 
changes  in the factors:
\begin{equation}
b_{i_k} \longmapsto b_{i_k}'= b_{i_k} \Delta_{i_k} \;\;\;\;\;\;\; k=1,2,...,N_c
\label{eq:smallbprime}
\end{equation}
with $\left\{ i_1 < i_2 < .... < i_{N_c} \right\}$.
In the case only one $b_{i_1}$ factor is changed ($N_c=1$) the 
determinant ratio can be easily obtained as:
\begin{eqnarray}
R&=&\frac{\det(1+B_L ... B_2 b_{N_b}...b_{i_1}\Delta_{i_1}...b_1)}
{\det(1+B_L ... B_2 B_1)} \\
&=& \det\left[1+(1-\overline{g}_{1})(\Delta_{i_1}-1)\right] \label{eq:ratioBSS}
\end{eqnarray}
where $\overline{g}_1=(1+b_{i_1}...b_1B_L ... B_2 b_{N_b}...b_{i_1})^{-1}$ 
is a modified electron Green Function and is obtained from $g$ as:
\begin{equation}
\overline{g}_1=(b_{i_1 -1}...b_1) g (b_{i_1 -1}...b_1)^{-1}
\end{equation}

The matrix $(\Delta_{j}-1)$ is symmetric and has only four matrix elements
different from zero, as can be seen from Eq. (\ref{eq:smallb},\ref{eq:smallbprime}), so that the products in Eq. (\ref{eq:ratioBSS}) can be performed in order 
$N n$ operation.

So far it is known how to calculate the determinant as long as a single
bond is changed. In the more complicate case of several bonds, the 
problem can be reduced to this simpler one by noting that the determinant
is expressed as:

\begin{equation}
R=R_{N_c,N_c-1} R_{N_c-1,N_c-2}...R_{1,0}
\end{equation}
and $R_{j,j-1}$ is the ratio between two determinants having changed only
the first $j$ and $j-1$ bonds respectively, 
\begin{equation}
R_{j,j-1}=\frac{\det(1+B_L ... B_2 b_{N_b}...b_{i_j}\Delta_{i_j}...b_{i_1}\Delta_{i_1}...b_1)}{\det(1+B_L ... B_2 b_{N_b}...b_{i_{j-1}}\Delta_{i_{j-1}}...b_{i_1}\Delta_{i_1}...b_1)}
\end{equation}
Each of these $N_c$ determinant ratios is given by Eq. 
(\ref{eq:ratioBSS}) with the Green function $g$ replaced by the new one
\begin{equation}
\overline{g}_{j-1}=(1+b_{i_{j}-1}...b'_{i_{j-1}}...b'_{i_1}...b'_1B_L...B_2 b_L...b_{i_{j}})^{-1}	
\end{equation}
which has only the first $j-1$ bonds updated.

Once the determinant $R_{j,j-1}$ has been obtained, it is necessary to
update the Green function $\overline{g}_{j-1}$ to the new one 
$\overline{g}_{j}$ which will be used to evaluate $R_{j+1,j}$.
This update is done in two steps and requires the knowledge of
$\overline{g}_{j}^{-1}$ so that the function $g^{-1}$ has to be bookkeeped
during the simulation.

The first step is to define the new Green function $\tilde{g}_{j}$ as:
\begin{equation}
\tilde{g}_{j}=(1+b_{i_{j}-1}...b'_{i_{j-1}}...b'_{i_1}...b'_1B_L...B_2 b_L...b'_{i_{j}})^{-1}
\end{equation}
$\tilde{g}_{j}$ differs from $\overline{g}_{j-1}$ only by the substitution 
$b_{i_j} \to b'_{i_{j}}$. It can be obtained using the Green function updating
in the simpler case of a single bond change \cite{Scalapino}, namely
\begin{equation}
\tilde{g}_j=[\overline{g}^{-1}_{j-1} + (\overline{g}^{-1}_{j-1}-1)(\Delta_{i_{j}}-1)]^{-1} \label{eq:updateoff1}
\end{equation}
The matrix $A=(\overline{g}^{-1}_{j-1}-1)(\Delta_{i_{j}}-1)$ is zero everywhere
a part from two columns. As a consequence, Eq. (\ref{eq:updateoff1}) can be
efficiently performed with the Shermann-Morrison formula \cite{numerical}
applied to $\overline{g}^{-1}_{j-1}$ so that the calculation of 
$\tilde{g}_{j}$  involves order $(N n)^2$ operations.

The second step is then to obtain from $\tilde{g}_j$ the Green function
$\overline{g}_j$ as follows:
\begin{equation}
\overline{g}_{j}=(b_{i_{j+1}-1}...b_{i_{j}+1}b'_{i_{j}})\tilde{g}_j
(b_{i_{j+1}-1}...b_{i_{j}+1}b'_{i_{j}})^{-1}\label{eq:updateoff2}
\end{equation}
Once $\overline{g}_j $is known it is clearly possible to obtain
$R_{j+1,j}$ following the same steps we have outlined before.

For a given Trotter slice and a given phonon coordinate the algorithm can 
be summarized as follows:

\begin{enumerate}

\item Displace coordinate $R_{i}\to R'_{i}$ and identify
the bonds ${i_1 < i_2 < ... < i_{N_c}}$ which will be
affected by the atomic displacement. 

\item Set $\tilde{g}_0=g$ and $\tilde{g}_0^{-1}=g^{-1}$, 
compute $\overline{g}_0$ and $\overline{g}_0^{-1}$ using 
Eq. (\ref{eq:updateoff2}) and the similar one for 
$\overline{g}_0^{-1}$.

\item Perform loop $j=1,...,N_c$ over the previously identified bonds.

\item Calculate the matrix $\Delta_{i_{j}}$

\item Calculate $R_{j,j-1}$ using $\overline{g}_{j-1}$ and
Eq. (\ref{eq:ratioBSS}).

\item Update $\overline{g}_{j-1}\to \overline{g}_{j}$ and
$\overline{g}_{j-1}^{-1} \to \overline{g}_{j}^{-1}$ using
Eqs. (\ref{eq:updateoff1},\ref{eq:updateoff2}).

\item End loop over $j$.

\item Compute R and check if the proposal move is accepted.

\item If the proposal is accepted update 
$\overline{g}_{N_c-1}\to \tilde{g}_{N_c}$ from  Eq. (\ref{eq:updateoff1}).

\end{enumerate}

After the proposed displacement for atom $i$ has been accepted by the 
Metropolis condition, the most straightforward way to proceed 
would be to obtain the new Green function $g'$
(eq. \ref{eq:green}), with all the $b_{\eta}$ factors updated, as
\begin{equation}
g'=(b_{i_{N_c}}...b_1)^{-1}\tilde{g}_{N_c}(b_{i_{N_c}}...b_1)\label{eq:gback}
\end{equation}	
and then from step $2$ of the algorithm obtain the new $\overline{g}_0'$
 for the  atom $j=i+1$. 
Note anyway that these two steps can be efficiently condensed in
one if a particular order for the sites is chosen. If the sites are ordered
in such a way that $i_1$ increase monotonically with $i$, e.g.
$\left\{1_1 < 2_1 < ... < N_{c_{1}}\right\}$, then Eq. (\ref{eq:gback})
becomes:
\begin{equation}
\tilde{g}_0'=(b_{i_{N_c}}...b_{j_1})^{-1}\tilde{g}_{N_c}
(b_{i_{N_c}}...b_{j_1})
\end{equation}
involving $2j_{i_1}$ products by $b_{\eta}$ factors less than the most 
straightforward procedure.

\section{Loss of momentum conservation}\label{sec:da}
At large $T$ the phonon vibrations become very large. In the semiclassical
treatment of the phonons, this tends to destroy the periodicity
and therefore it tends to violate momentum conservation within the
electronic system. Below we test how this violation increases with $T$ in
the TM model using a HI coupling. Qualitatively similar results
are, however, obtained also in the other models.                           
We first calculate the states of the Hamiltonian at $T=0$. The system
is then perfectly periodic and all the states $|n{\bf k},T=0\rangle$ 
can be labelled by a wave vector ${\bf k}$ and a band index $n$.  
We use a unit cell with six Nb atoms and the band index therefore
runs over 30 states. Next the states at a finite $T$ are calculated. 
These states $|l,T\rangle$ are labelled by an index $l$. These states 
can be expanded in the complete set of $T=0$ states 
\begin{equation}\label{eq:l1}
|l,T\rangle=\sum_{{\bf k}n} |n{\bf k},T=0\rangle\langle n{\bf k},T=0|
l,T\rangle,
\end{equation}
For a given state we determine the amount of ${\bf k}$-character 
\begin{equation}\label{eq:l2}
c_{\bf k}(l)=\sum_{n} |\langle n{\bf k},T=0|l,T\rangle|^2.
\end{equation}
or the amount of mixing with states having the band index $n$
\begin{equation}\label{eq:l2a}
c_n(l)=\sum_{{\bf k}} |\langle n{\bf k},T=0|l,T\rangle|^2.
\end{equation}
From normalization it follows that $\sum_{\bf k} c^{(l)}_{\bf k}=1$
and $\sum_{n} c^{(l)}_{n}=1$.
We define
\begin{equation}\label{eq:l3}
\Delta_k(l)=n_{\bf k}\sum_{\bf k} \lbrack c^{(l)}_{\bf k}\rbrack^2,
\end{equation}
and
\begin{equation}\label{eq:l3a}
\Delta_n(l)=30\sum_{n} \lbrack c^{(l)}_{n}\rbrack^2,
\end{equation}
where $n_{\bf k}$ is the number of allowed ${\bf k}$-vectors
and 30 is the number of band index.
If a weight of a given state $l$ is equally distributed over $n_{\bf k}/m$
different ${\bf k}$-vectors, $\Delta_k(l)=m$. In particular, if all effects 
of periodicity are lost, we expect that $\Delta_k(l)=1$, since we then expect
all $n_{\bf k}$ ${\bf k}$-components to have equal weight ($m=1)$.
On the other hand, if a state
contains only one ${\bf k}$-vector, $\Delta_k(l)=n_{\bf k}$. Typically in the
periodic system, several states with different ${\bf k}$-vectors
are degenerate, e.g., states with ${\bf k}$ and -${\bf k}$ may be degenerate. 
Even at a very small amount of disorder, a state of the disordered system
is then typically a linear combination of states with several different
${\bf k}$-vectors, and $\Delta_k(l)$ is reduced correspondingly.
We consider a super cell with periodic boundary conditions. The value
of $n_{\bf k}$ then depends on the size of the super cell. For a 
given amount of disorder, we expect that a given state will contain
${\bf k}$-vectors from a certain fraction of the Brillouin zone.
The number of ${\bf k}$-vectors increases with the size of the
super cell. However, $m$ introduced above should stay roughly constant.
Thus we find that definition (\ref{eq:l3}) gives results which are
rather independent of the super cell size for values of $T$ which are not
very small. On the other hand, for $T\approx 0$, this definition gives
results which grow roughly linearly with $n_{\bf k}$. The definition is,
however, sensible for the range of $T$ of interest here.
In a similar way it follows that $\Delta_n(l)=1$ if the conservation of
the band indices is completely lost.

\begin{figure}[bt]
\centerline{
\rotatebox{-90}{\resizebox{!}{3.0in}{\includegraphics{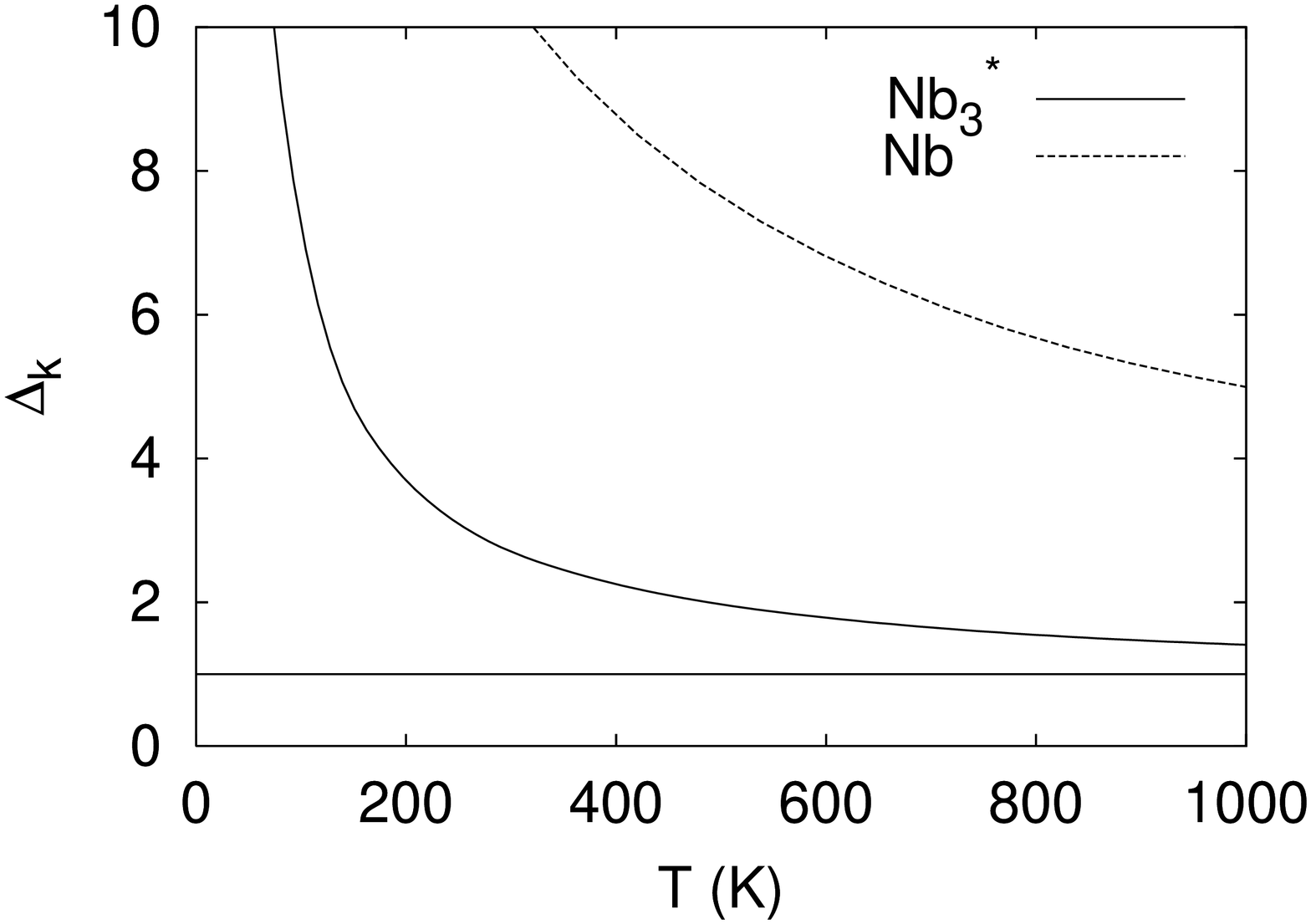}}}}
\caption[]{\label{fig:per}The quantity $\Delta_k$ in Eq. (\ref{eq:l4})
for Nb$_3^{\ast}$ and Nb as a function of $T$ for $n_{\bf k}=256$  
allowed ${\bf k}$-vectors. $\Delta_k$ measures the loss of periodicity. 
The horizontal line ($\Delta_k=1$) represents complete loss of 
periodicity. The figure illustrates the rapid loss of periodicity and
momentum conservation for the Nb$_3^{\ast}$ model, while this
loss happens more slowly for Nb.}
\end{figure}

We average over all states              
\begin{equation}\label{eq:l4}
\Delta_i=\sum_l\Delta_i(l)/(Nn)\hskip0.5cm i=k \ {\rm or} \ n.
\end{equation}
Fig. \ref{fig:per} shows $\Delta_k$   
for Nb$_3^{\ast}$ and Nb, where $\Delta$ is an average over 
$\Delta(l)$. The line $\Delta=1$, corresponding to a complete 
loss of periodicity, is also shown. The figure illustrates that 
for Nb$_3^{\ast}$ much of the periodicity is lost already for 
$T\sim 200-300$ K. For Nb this happens at higher $T$, but 
also in this case periodicity is lost fairly quickly.   

The rapid loss of periodicity for Nb$_3^{\ast}$ can be related to
the many flat bands. This means that there are states with all       
${\bf k}$-values within a rather small energy range. Then only
a small perturbation is needed to mix all these different 
${\bf k}$-values, implying a loss of momentum conservation.  

\begin{figure}[bt]
\centerline{
\rotatebox{-90}{\resizebox{!}{3.0in}{\includegraphics{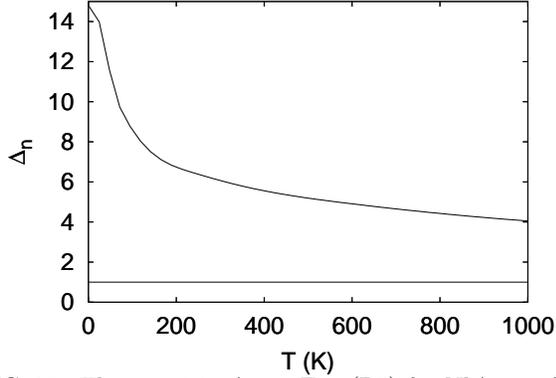}}}}
\caption[]{\label{fig:pern}The quantity $\Delta_n$ in Eq. (\ref{eq:l4})
for Nb$_3^{\ast}$ as a function of $T$ for $n_{\bf k}=256$  
allowed ${\bf k}$-vectors. $\Delta_n$ measures how the conservation of 
the band index is lost, with the horizontal line ($\Delta_n=1$) showing    
a complete loss. The figure illustrates how the meaning of the band indices
is lost relatively rapidly for the Nb$_3^{\ast}$ model.}
\end{figure}

In a similar way, Fig. \ref{fig:pern} shows that the meaning of
the band indices is lost relatively quickly for Nn$_3^{\ast}$
as $T$ is increased. This means that the meaning of intraband and 
interband transitions start to loose their meaning. 

\section{Constant current matrix elements}\label{sec:db}

\begin{figure}
\hskip-0.1in
\rotatebox{-90}{\resizebox{2.50in}{!}{\includegraphics{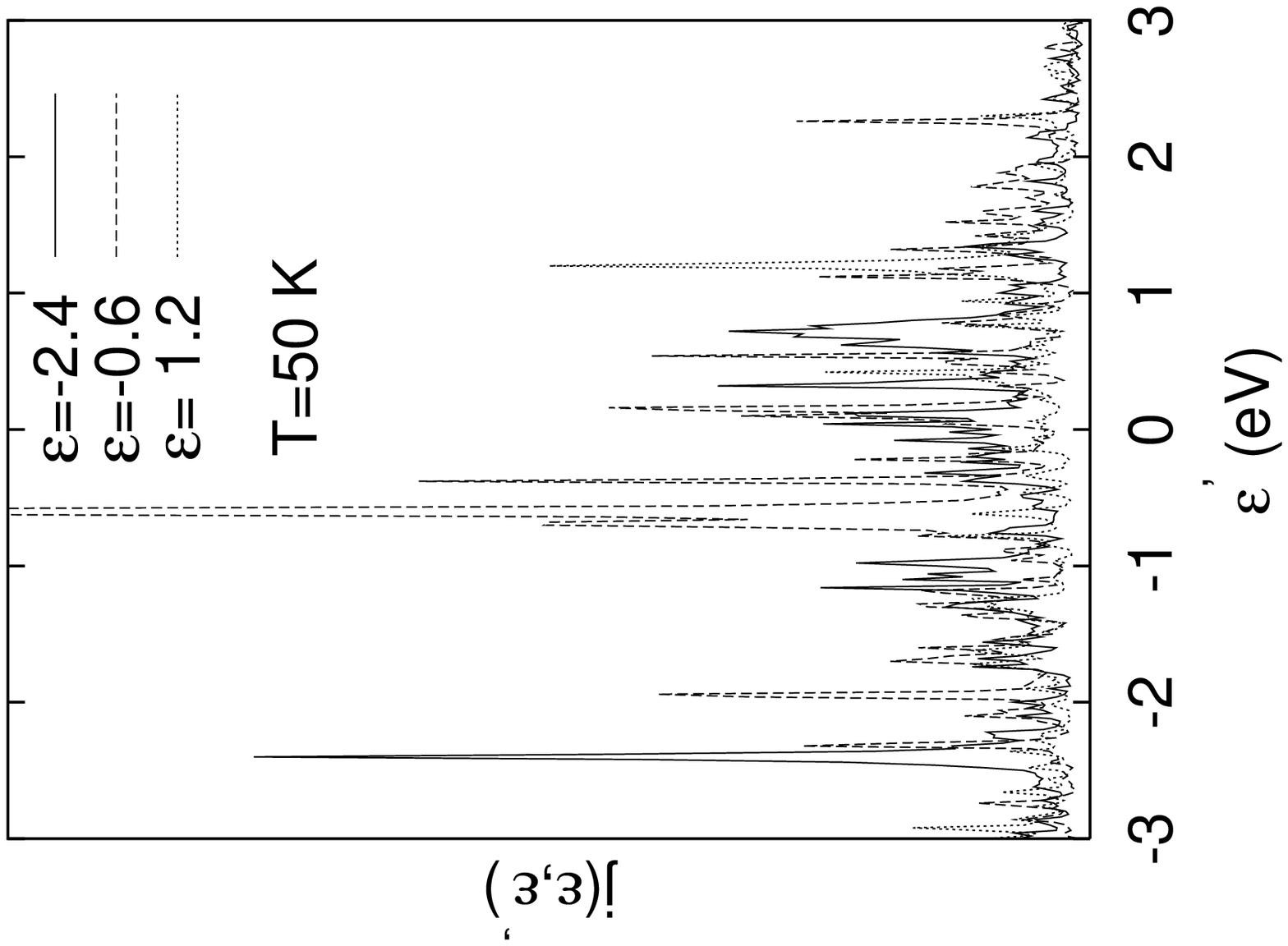}}}
\hskip-0.3cm
\rotatebox{-90}{\resizebox{2.50in}{!}{\includegraphics{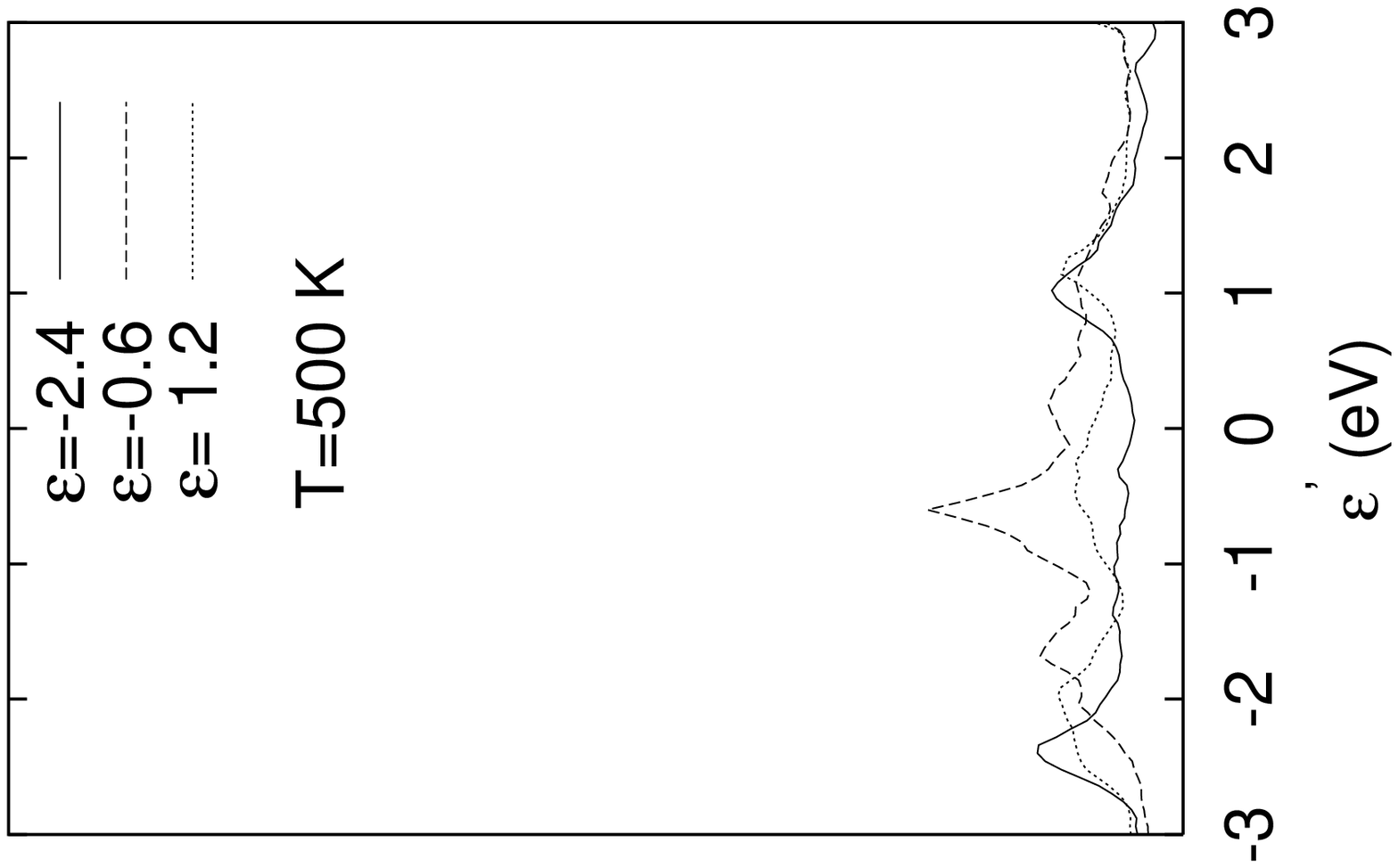}}}
\caption[]{\label{fig:J}Average $j(\varepsilon,\varepsilon^{'})$
of the current matrix elements over states with similar energies 
(Eq. (\ref{eq:d2})) for Nb$_3^{\ast}$. The units are arbitrary.
The figure illustrates that for $T=0.0043$ eV=50 K $j(\varepsilon,
\varepsilon^{'})$  varies strongly with the energies while for
$T=0.043$ eV=500 K this variation is much less pronounced.}
\end{figure}

In view of the rapid loss of momentum conservation, illustrated 
in Appendix \ref{sec:da}, it is interesting to consider the limit 
where momentum conservation is completely lossed due to the disorder.
This is the opposite limit to the traditional Bloch-Boltzmann treatment,
where the scattering is assumed to be so small that ${\bf k}$ is a 
useful quantum number. In the complete disorder limit studied here,
all states are coupled to all states via the current operator. 
The calculations for the Nb$_3^{\ast}$ model show that these assumptions,
taken literally, are not satisfied. We note, however, 
that the expression in Eq. (\ref{eq:me3}) for the optical conductivity 
can be rewritten as
\begin{eqnarray}\label{eq:d1} 
&&\sigma(\omega)={2\pi n^2\over N\Omega \omega}\int d\varepsilon N(\varepsilon)
\int d\varepsilon^{'} N(\varepsilon^{'})j(\varepsilon,\varepsilon^{'})
\nonumber \\
&&\times \lbrack f(\varepsilon)-f(\varepsilon^{'})\rbrack
\delta(\hbar \omega-\varepsilon^{'}+\varepsilon),
\end{eqnarray}
where 
\begin{equation}\label{eq:d2}
j(\varepsilon,\varepsilon^{'})={1\over n^2 N(\varepsilon)N(\varepsilon^{'})}
\sum_{ll^{'}}|\langle l|j_x|l^{'}\rangle|^2
L(\varepsilon-\varepsilon_l)L(\varepsilon^{'}-\varepsilon_l^{'}),
\end{equation}
$N(\varepsilon)$ is the density of states per atom, orbital and spin
and $n$ is the orbital degeneracy.
$L(\varepsilon)=(\gamma/\pi)/(\varepsilon^2+\gamma^2)$ is a 
Lorentzian. The function $j(\varepsilon,\varepsilon^{'})$ is shown 
in Fig. \ref{fig:J} for two values of $T$, using the broadening 
$\gamma=0.01$ eV.
The figure illustrates that the function $j(\varepsilon,\varepsilon^{'})$
has only a moderate dependence on the energies for $T=0.043$ eV=500 K. 
We therefore now work out the consequences of assuming that the 
matrix elements of the current can be replaced by their average. 

This average is defined as
\begin{equation}\label{eq:d3}
j_{av}^2={1\over (Nn)^2} \sum_{ll^{'}}|\langle l|\hat j_x|l^{'}\rangle|^2,
\end{equation}
where $|i\rangle$ are the $Nn$ eigenstates of the Hamiltonian.
The expression (\ref{eq:me3}) for the optical
conductivity can then be written as
\begin{equation}\label{eq:d2a}
\sigma(\omega)={2\pi\over N\Omega}j_{av}^2\sum_l^{occ}\sum_{l^{'}}^{unocc}
{1\over \omega}\delta(\hbar\omega-\varepsilon_{l^{'}}+\varepsilon_l),
\end{equation}
where we have considered $\omega>0$ and assumed that $T<<W$ so that
we can replace the Fermi functions by $\Theta$-functions.
Fig. \ref{fig:dsigma}a compares the actually calculated $\sigma(\omega)$
with the result of (\ref{eq:d2a}), assuming a semi-elliptical DOS 
(Eq. (\ref{eq:d2a})). The good agreement for large $T$
gives further justification for the assumptions behind Eq. (\ref{eq:d2a}).
This gives         
\begin{equation}\label{eq:d2b}
\sigma(\omega=0)={2\pi N n^2 \hbar\over \Omega}j_{av}^2N(\mu)^2,
\end{equation}
where $\mu$ is the chemical potential. We then need to find a relation
between $j_{av}$ and $N(\mu)$, which is obtained from 
charge and current conservation. We first rewrite $j_{av}$ as 
\begin{equation}\label{eq:d2c}
j_{av}^2
={1\over (Nn)^2} \sum_{\nu\mu}|\langle \nu|\hat j_x|\mu\rangle|^2,
\end{equation}
where $|\nu \rangle$ is a basis state in a local representation. We 
then use the charge and current conservation in Eq. (\ref{eq:m4}), 
relating the current and hopping matrix elements. This gives
\begin{equation}\label{eq:d4}
\sum_{\alpha=x,y,z}|\langle \nu|\hat j_{\alpha}|\mu\rangle|^2=
{e^2d^2 \over \hbar^2} t_{\nu\mu}^2.
\end{equation} 
and for an isotropic system                        
\begin{equation}\label{eq:d5}
j_{av}^2={1\over (Nn)^2}{e^2d^2\over 3\hbar^2}\sum_{\nu\mu} t_{\nu\mu}^2.
\end{equation} 
To relate $j_{av}^2$ to $N(\varepsilon)$, we introduce the 
second moment
\begin{equation}\label{eq:d6}
\langle \varepsilon^2 \rangle = \int_{-\infty}^{\infty} N(\varepsilon)
\varepsilon^2 d\varepsilon,
\end{equation}
where $N(\varepsilon)$ is normalized to unity.               
This quantity can be related to the 
hopping integrals
\begin{equation}\label{eq:d7}
n\langle \varepsilon^2 \rangle ={1\over N}\sum_{\nu\mu}t_{\nu\mu}^2.
\end{equation}
We assume a specific form for $N(\varepsilon)$, calculate
$\langle \varepsilon^2 \rangle$ for this form and then relate it to
$N(\mu)$. Table \ref{table:d1c} shows results for different shapes
of the DOS.  The table illustrates that there is not
a drastic dependence on the shape of $N(\varepsilon)$. In the
following, we focus the semi-elliptical DOS, which is probably the
most realistic one of the three cases considered.

\begin{minipage}{3.375in}                         
\begin{table}
\caption[]{\label{table:d1c}The quantity $\langle \varepsilon^2 \rangle
N(\mu)^2$ for a constant (Eq. (\ref{eq:d8})), a Gaussian 
(Eq. (\ref{eq:d9})) and a semi-elliptical (Eq. (\ref{eq:d10}))
density of states (DOS) and for half-filling.}               
\begin{tabular}{cccc}
 & Constant & Gaussian & Semi-elliptical   \\
$\langle \varepsilon^2 \rangle N(\mu)^2$ & ${1\over 12}=0.083$ &
${1\over 2\pi}=0.159$ & ${1\over \pi^2}=0.101 $   \\

\end{tabular}
\end{table}
\end{minipage}

Expressing $\sum t_{\nu\mu}$ in terms of $\langle \varepsilon^2 \rangle$ in
Eq. (\ref{eq:d5}), we can rewrite Eq. (\ref{eq:d2b}) as
\begin{equation}\label{eq:d11}
\sigma(0)={2\pi n\over 3}{d^3\over \Omega}
\langle \varepsilon^2 \rangle N(\mu)^2 {e^2\over d \hbar},
\end{equation}
where $\Omega/d^3$ is shown in Table \ref{table:d2}. 
The quantity $e^2/(\hbar d)$ has the unit of conductivity
and Eq. (\ref{eq:d11}) can be rewritten as
\begin{equation}\label{eq:d12}
\rho={1\over \sigma(0)}=19.7 {\Omega/d^3 \over 
\langle \varepsilon^2\rangle N(\mu)^2}{d\over n} \ \mu\Omega{\rm cm},
\end{equation}
where $d$ is now expressed in \AA.
As seen in Tables \ref{table:d1c} and \ref{table:d2}, 
$\langle \varepsilon^2\rangle N(\mu)^2 \sim 0.1$ and $\Omega/d^3\sim 1$. 
For a transition metal, we may use $d\sim 3$ \AA \ and $n=5$, 
which leads to $\rho\sim 100 \ \mu\Omega$cm. Such a resistivity
is indeed typical for the saturation resistivity of a transition metal
compound.

\section{Derivation of the \lowercase{f}-sum rule}\label{sec:f}

In this appendix we derive the f-sum rule, essentially following 
Maldague.\cite{Maldague}
We introduce the position operator
\begin{equation}\label{eq:af1}
\hat R_x=\sum_{\nu\sigma}R_x^{\nu}\psi_{\nu\sigma}^{\dagger}
\psi_{\nu\sigma}.
\end{equation}
The current operator can then be expressed as 
\begin{equation}\label{eq:af2}
\hat j_x={ie\over \hbar}[\hat R_x,H].
\end{equation}
For $\omega>0$, the optical conductivity is written as
\begin{equation}\label{eq:af3}
\sigma(\omega)={\pi\hbar\over N\Omega}\sum_{n}|\langle n |\hat j_x|0\rangle|^2
{\delta(\hbar|\omega|-E_n+E_0)\over E_n-E_0}
\end{equation}
where $|n\rangle$ is a many-body state with the energy $E_n$.
By inserting Eq. (\ref{eq:af2}) in one of the two matrix elements of
$\hat j_x$, one obtains
\begin{equation}\label{eq:af4}
{2\over \pi}\int_0^{\infty} \sigma(\omega)d \omega= 
{e^2\over N\Omega \hbar^2}\langle 0|[[H,\hat R_x],\hat R_x]|0\rangle.
\end{equation}
Performing the commutators, we find
\begin{equation}\label{eq:af5}
\sum_{\alpha}[[H,\hat R_{\alpha}],\hat R_{\alpha}]=\sum_{\nu\mu}
d_{\nu\mu}^2 t_{\mu\nu}\psi^{\dagger}_{\nu\sigma}\psi_{\mu\sigma},
\end{equation}
where $d_{\nu\mu}$ is the distance between the sites with the orbitals
$\nu$ and $\mu$ and $\alpha$ labels the coordinate. This result is
true for noninteracting systems as well as interacting systems of 
certain types, e.g., with an on-site Hubbard interaction. We now assume
only nearest neighbor hopping, replacing $d_{\nu\mu}$ by $d$. Furthermore,
we assume the system to be isotropic, so that all directions $\alpha$ 
are equivalent. For a three-dimensional system, the commutator on the
right hand side of Eq. (\ref{eq:af4}) is then one third of the result in
Eq. (\ref{eq:af5}). This gives
\begin{equation}\label{eq:af6}
{2\over \pi}\int_0^{\infty} \sigma(\omega)d \omega=
-{1\over 3}{d^2e^2\over N\Omega\hbar^2}\langle 0|T_K|0\rangle,
\end{equation}
where $T_K$ is the kinetic energy.  For a two-dimensional system the 
factor 3 in the denominator is replaced by a factor 2. This result 
can also be generalized to a finite temperature. In the case of the TM
model, however, the atomic separations cannot be treated as constants,
since they vary as the phonons are excited. The coordinates in 
Eq. (\ref{eq:af4}) can then not be taken outside the average 
$\langle ... \rangle$. We can, nevertheless, recover an expression
like Eq. (\ref{eq:af6}) by defining an appropriate average separation
$d(T)$.

\end{multicols}

\begin{thebibliography}{*}
\bibitem{scatteringcenters}For La$_{1.96}$Sr$_{0.04}$CuO$_4$ 
we have assumed a cylindrical Fermi surface with the height 
$2\pi/c$, where $c$ is the distance between the CuO$_2$ planes. 
The mean free path was assumed to be equal to the Cu-Cu distance in 
the planes. For C$_{60}$ we assumed the mean free path to be equal to 
the separation of the C$_{60}$ molecules.
\end{thebibliography}
\end{document}